\documentclass{article}
\usepackage{amsmath}
\usepackage[english]{babel}
\usepackage{latexsym}
\usepackage{amssymb}
\usepackage{amscd}

\numberwithin{equation}{section}
\newcommand{\bdm}{\begin{displaymath}}
\newcommand{\edm}{\end{displaymath}}
\newcommand{\bdn}{\begin{eqnarray}}
\newcommand{\edn}{\end{eqnarray}}
\newcommand{\bay}{\begin{array}{c}}
\newcommand{\eay}{\end{array}}
\newcommand{\ben}{\begin{enumerate}}
\newcommand{\een}{\end{enumerate}}
\newcommand{\martin}{\begin{equation}}
\newcommand{\sileno}{\end{equation}}
\newtheorem{lem}{Lemma}[section]
\newtheorem{teo}{Theorem}[section]
\newtheorem{pro}{Proposition}[section]

\newtheorem{cor}{Corollary}[section]

\title{Rotating Singular Perturbations of the Laplacian}
\author{Michele Correggi\footnote{E-mail address: \texttt{correggi@sissa.it}}	\\	\small{International School for Advanced Studies SISSA/ISAS,}	\\	\small{Trieste, Italy}	\\	\mbox{}	\\	Gianfausto Dell'Antonio\footnote{E-mail address: \texttt{gianfa@sissa.it}}	\\	\small{Centro Linceo Interdisciplinare}\footnote{On leave from Dipartimento di Matematica, Universit\`{a} di Roma, ``La Sapienza'', Italy.},	\\	\small{Roma, Italy}}

\begin{document}

\maketitle

\begin{abstract}
	We study a system of a quantum particle interacting with a singular time-dependent uniformly rotating potential in 2 and 3 dimensions: in particular we consider an interaction with support on a point (rotating point interaction) and on a set of codimension 1 (rotating blade). We prove the existence of the Hamiltonians of such systems as suitable self-adjoint operators and we give an explicit expression for the unitary dynamics. Moreover we analyze the asymptotic limit of large angular velocity and we prove strong convergence of the time-dependent propagator to some one-parameter unitary group as \( \omega \rightarrow \infty \).
\end{abstract}

\section{Introduction}

In this paper we shall study systems defined by formal time-dependent Schr\"{o}dinger operators on \( L^2(\mathbb{R}^n) \), \( n = 2,3 \)
	\martin
		\label{Ham1}
		H(t) = H_0 + V_t = - \Delta + V_t
	\end{equation}
with uniformly rotating potentials
	\martin
		\label{Pot1}
		V_t(\vec{x}) = V(\mathcal{R}^{-1}(t) \: \vec{x})
	\end{equation}
where \( V \) is a singular potential (e.g. \( V(\vec{x}) = \delta(\vec{x} - \vec{y}_0) \)) and \( \mathcal{R}(t) \) a rotation on the \( x,y-\)plane with period \( 2 \pi / \omega \):
	\bdm
		\mathcal{R}(t)	= \left(
		\begin{array}{ccc}
			\cos(\omega t)	&	-\sin(\omega t)	&	0	\\
			\sin(\omega t)	&	\cos(\omega t)	&	0	\\
			0		&			&	1	\\
		\end{array}
		\right)
	\edm
Regular rotating potentials were studied by Enss et al. \cite{Enss1} in order to extract information about the scattering of a quantum particle: indeed they considered a class of potentials such that the kinetic energy of the system remains bounded on the range of wave operators and they proved existence and completeness of the wave operators. 
\newline
Our purpose is to define in a rigorous way the time-dependent Hamiltonians (\ref{Ham1}) when the potential has a more singular behavior: we shall study rotating point perturbations\footnote{Point interactions were introduces for the first time in a rigorous way by Berezin and Faddeev in 1961 \cite{Bere1}. For general references about fixed and time-dependent point interactions see \cite{Albe1, Dell1, Dell2, Figa1}.} of the Laplacian in 2 and 3 dimensions and rotating blades, namely rotating singular potentials supported over a set of codimension 1 (a segment in 2 dimensions and an half-disk in 3 dimensions respectively). 
\newline
As pointed out by Enss et al., the uniformly rotating Hamiltonians can be studied in a simpler way than general time-dependent operators, indeed, considering the time evolution \( U_{\mathrm{rot}}(t,s) \) of the system in a uniformly rotating frame around the \(z-\)axis, it is easy to see that the following relation with the time evolution in the inertial frame \( U_{\mathrm{inert}}(t,s) \) holds
	\martin
		U_{\mathrm{inert}}(t,s) = R(t) \: U_{\mathrm{rot}}(t-s) \: R^{\dagger}(s)
	\end{equation}
where \( R(t) \Psi(\vec{x}) = \Psi(\mathcal{R}(t)^{-1} \: \vec{x}) \) and \( U_{\mathrm{rot}}(t,s) = U_{\mathrm{rot}}(t-s) \) is the one-parameter unitary group 
	\martin
		U_{\mathrm{rot}}(t-s) = e^{-i K (t-s)}
	\end{equation}
with a time-independent generator \( K \), formally defined in the following way
	\martin
		K = H_0 - \omega J + V
	\end{equation}
Here \( J \) stands for the third component of the angular momentum and \( V \) is the time-independent potential (\ref{Pot1}).
\newline
Using this trick we shall define the previous time-dependent Hamiltonians considering the corresponding formal time-independent generators in the rotating frame and studying their self-adjoint extensions. 
\newline
The last goal of this work will be the analysis of the asymptotic limit of the systems when the angular velocity \( \omega \rightarrow \infty \): by means of the explicit expression of resolvents of singular perturbations of the Laplacian, we shall prove convergence in strong sense of \( U_{\mathrm{inert}}(t,s) \) to some one-parameter unitary group \( U_{\mathrm{asympt}}(t-s) \) with time-independent generator \( H_{\mathrm{asympt}} \). Moreover we shall see that, for point interactions, \( H_{\mathrm{asympt}} \) is the Laplacian with singular perturbation on a circle, while the asymptotic limit of the rotating blade is simply a regular potential supported on a compact set. The same study was performed by Enss et al. \cite{Enss2} for regular rotating potentials.

\section{The Rotating Point Interaction in 3D}

\subsection{The Hamiltonian}

The system we shall study is defined by the formal time-dependent Hamiltonian
	\martin
		H(t) = H_0 + a \: \delta^{(3)}(\vec{x} - \vec{y}(t))
	\end{equation}
where \( \vec{y}(t) = \mathcal{R}(t) \vec{y}_0 \). 
\newline
According to the previous scheme, the formal generator of time evolution in the uniformly rotating frame (with angular velocity \( \omega \)) is given by
	\bdm
		K = H_0 - \omega J + a \: \delta^{(3)}(\vec{x} - \vec{y}_0)
	\edm
Therefore the Hamiltonian of the system is a self-adjoint extension of the operator
	\bdm
		K_{y_0} = H_{\omega}
	\edm
	\bdm
		\mathcal{D}(K_{y_0}) = C^{\infty}_0 (\mathbb{R}^3 - \{\vec{y}_0\})
	\edm
The operator \( K_{y_0} \) is symmetric and then closable; let \( \dot{K}_{y_0} \) be its closure, with domain \( \mathcal{D}(\dot{K}_{y_0}) \).
\newline
The function
	\martin
		\label{Res1}
		\mathcal{G}_z(\vec{x},\vec{y}_0) = \int_0^{\infty} dk \sum_{l=0}^{\infty} \sum_{m = -l}^{l} \frac{1}{k^2 - m \omega - z} \: \varphi^*_{klm} (\vec{y}_0) \: \varphi_{klm}(\vec{x}) 
	\end{equation}
for \( \vec{x} \in \mathbb{R}^3 - \{ \vec{y}_0 \} \) and \( z \in \mathbb{C} - \mathbb{R} \), is the unique solution of
	\bdm
		\dot{K}_{y_0}^* \Psi_z(\vec{x}) = z \Psi_z(\vec{x})
	\edm
with \( \Psi \in \mathcal{D}(\dot{K}_{y_0}^*) \) (see Proposition \ref{Gre1}).
\newline
The operator \( \dot{K}_{y_0} \) has then deficiency indexes \( (1,1) \) and its self-adjoint extensions are given by the one-parameter family of operators \( K_{\alpha,y_0} \), \( \alpha \in [0,2\pi) \):
	\martin
		\mathcal{D}(K_{\alpha,y_0}) = \{ f + c \mathcal{G}_+ + c e^{i\alpha} \mathcal{G}_- \: | \: g \in \mathcal{D}(\dot{K}_{y_0}), c \in \mathbb{C} \}
	\end{equation}
	\martin
		K_{\alpha,y_0} ( f + c \mathcal{G}_+ + c e^{i\alpha} \mathcal{G}_- ) = \dot{K}_{y_0} g + i c \mathcal{G}_+ - i c e^{i\alpha} \mathcal{G}_-
	\end{equation}
where
	\bdm
		\mathcal{G}_{\pm}(\vec{x}) = \mathcal{G}_{\pm i} (\vec{x}, \vec{y}_0) = \int_0^{\infty} dk \sum_{l=0}^{\infty} \sum_{m = -l}^{l} \frac{1}{k^2 - m \omega \mp i} \: \varphi^*_{klm} (\vec{y}_0) \: \varphi_{klm}(\vec{x}) 
	\edm
for \( \vec{x} \in \mathbb{R}^3 - \{\vec{y}_0\} \). 
\newline
Moreover the self-adjoint extension \( K_{\pi,y_0} \) corresponds to the ``free'' Hamiltonian \( \dot{H}_{\omega} \): indeed, if \( \Psi \in \mathcal{D}(K_{\pi,y_0}) \), 
	\bdm
		\Psi = f + c ( \mathcal{G}_+ - \mathcal{G}_- )
	\edm 
and the difference \( \mathcal{G}_+ - \mathcal{G}_- \) is a continuous function at \( \vec{x} = \vec{y}_0 \), which  belongs to the domain of \( H_{\omega} \), so that \( K_{\pi,y_0} \) becomes exactly the operator \( \dot{H}_{\omega} \). 
\newline
Using this result and applying the Krein's theory of self-adjoint extensions, it is easy to obtain the following
	
	\begin{teo}
		\label{Res2}
		The resolvent of \( K_{\alpha, y_0} \) has integral kernel given by
		\martin	
			(K_{\alpha,y_0} - z)^{-1} (\vec{x}, \vec{x}') = \mathcal{G}_z(\vec{x}, \vec{x}') + \lambda(z, \alpha) \mathcal{G}^*_{\bar{z}}(\vec{x}', \vec{y}_0) \mathcal{G}_z(\vec{x},\vec{y}_0)
		\end{equation}
with \( z \in \varrho(K_{\alpha,y_0}) \), \( \vec{x}, \vec{x}' \in \mathbb{R}^3 \), \( \vec{x} \neq \vec{x}' \), \( \vec{x}, \vec{x}' \neq \vec{y}_0 \) and
		\martin
		\label{lambda1}
			\frac{1}{\lambda(z, \alpha)} = \frac{1}{\lambda(-i, \alpha)} - (z+i) \big( \mathcal{G}_{\bar{z}} (\vec{x}) , \mathcal{G}_- (\vec{x}) \big)
		\end{equation}
		\martin
			\lambda(-i, \alpha) = \frac{1 + e^{i\alpha}}{2i \| \mathcal{G}_- (\vec{x}) \|^2}
		\end{equation}
	\end{teo}

	\emph{Proof:} 	Since \( \dot{K}_{y_0} \) is a densely defined, closed, symmetric operator with deficiency indexes \( (1,1) \), we can apply Krein's theory (cfr. \cite{Albe1, Posi1}) to classify all its self-adjoint extensions: from Krein's formula we immediately obtain
		\bdm
			(K_{\alpha, y_0} - z)^{-1} - (K_{\pi,y_0} - z)^{-1}  = \lambda(z, \alpha) \big( \mathcal{G}_{\bar{z}}(\vec{x}), \cdot \big) \mathcal{G}_z(\vec{x})
		\edm
		for \( z \in \varrho(K_{\alpha,y_0}) \cap \varrho(H_{\omega}) \). It follows that \( (K_{\alpha, y_0} - z)^{-1} \) has integral kernel given by
		\bdm
			(K_{\alpha, y_0} - z)^{-1} (\vec{x}, \vec{x}') = (\dot{H}_{\omega} - z)^{-1} (\vec{x}, \vec{x}') + \lambda(z, \alpha) \mathcal{G}^*_{\bar{z}}(\vec{x}', \vec{y}_0) \mathcal{G}_z(\vec{x},\vec{y}_0)
		\edm
		Moreover \( \lambda(z, \alpha) \) satisfies the following equation
		\bdm
			\frac{1}{\lambda(z, \alpha)} = \frac{1}{\lambda(z', \alpha)} - (z-z') \big( \mathcal{G}_{\bar{z}} (\vec{x}) , \mathcal{G}_{z'} (\vec{x}) \big)
		\edm
		The explicit expression of the factor \( \lambda(-i, \alpha) \) is given in the following Theorem.
		\begin{flushright} 
			\( \Box \)
		\end{flushright}
	
	\begin{teo}
		\label{Dom1}
		The domain \( \mathcal{D}(K_{\alpha, y_0}) \), \( \alpha \in [0,2\pi) \), consists of all elements \linebreak \( \Psi \in \mathbb{R}^3 \) which can be decomposed in the following way
		\bdm
			\Psi (\vec{x}) = \Phi_z(\vec{x}) + \lambda(z, \alpha) \Phi_z(\vec{y}_0) \mathcal{G}_z(\vec{x}, \vec{y}_0)
		\edm
		for \( \vec{x} \neq \vec{y}_0 \), \( \Phi_z \in \mathcal{D}(\dot{H}_{\omega}) \) and \( z \in \varrho(K_{\alpha, y_0}) \). The previous decomposition is unique and on every \( \Psi \) of this form 
		\bdm
			(K_{\alpha,y_0} - z) \Psi = (H_{\omega} -z) \Phi_z
		\edm
	\end{teo}

	\emph{Proof:} 	First of all we observe that functions belonging to \( \mathcal{D}(\dot{H}_{\omega}) \) are H\"{o}lder continuous with exponent smaller than \( 1/2 \) in every compact subset of \( \mathbb{R}^3 \). Indeed the domain of self-adjointness of \( \dot{H}_{\omega} \) contains functions in \( H^2_{\mathrm{loc}}(\mathbb{R}^3) \): on every compact set \( S \subset \mathbb{R}^3 \), the domain\footnote{The notation \( A^S \) denotes the restriction of the operator \( A \) to the Hilbert space \( L^2(S) \).} of \( H^S_0 \) is strictly contained on the domain of \( J^S \), since \( J^S \) is a bounded operator on \( \mathcal{D}(\dot{H}_0^S) = H^2(S) \), therefore \( \mathcal{D}(\dot{H}^S_{\omega}) = \mathcal{D}(\dot{H}^S_0) = H^2(S) \). Hence it makes sense to write \( \Phi(\vec{y}_0) \) for every \( \Phi \in \mathcal{D}(\dot{H}_{\omega}) \) and \( \vec{y}_0 \in \mathbb{R}^3 \).
		\newline
		Moreover
		\bdm
			\mathcal{D}(K_{\alpha ,y_0}) = (K_{\alpha,y_0} -z)^{-1} (\dot{H}_{\omega} - z) \mathcal{D}(\dot{H}_{\omega}) 
		\edm
		and the claim follows from the expression of the resolvent given in the previous Theorem \ref{Res1}. 
		\newline
		To prove the uniqueness of the decomposition let \( \Psi = 0 \), so that
		\bdm
			\Phi_z(\vec{x}) = - \frac{1 + e^{i\alpha}}{2i \| \mathcal{G}_- (\vec{x}) \|^2} \Phi_z(\vec{y}_0) \mathcal{G}_{z}(\vec{x})
		\edm
		but \( \Phi_z(\vec{x}) \) must be continuous at \( \vec{x} = \vec{y}_0 \): it follows that \( \Phi_z(\vec{y}_0) = 0 \) and then \( \Phi_z = 0 \).
		\newline
		Finally the last equality of the Theorem easily follows from
		\bdm
			(K_{\alpha, y_0} -z)^{-1} (\dot{H}_{\omega}-z) \Phi_z = \Phi_z + \lambda(z, \alpha) \big( \mathcal{G}_{\bar{z}}(\vec{x}) ,  (\dot{H}_{\omega}-z) \Phi_z (\vec{x}) \big)  \mathcal{G}_z = \Psi 
		\edm
		To find the explicit expression of \( \lambda(-i, \alpha) \) it is sufficient to study the behavior of functions in \( \mathcal{D}(K_{\alpha, y_0}) \) at \( \vec{y}_0 \). Let \( \Psi(\vec{x}) \in \mathcal{D}(K_{\alpha, y_0}) \),
		\bdm
			\Psi(\vec{x}) = f(\vec{x}) + c \mathcal{G}_+(\vec{x}) + c e^{i \alpha} \mathcal{G}_-(\vec{x})
		\edm
		with \( f \in \mathcal{D}(\dot{H}_{y_0}) \) and \( c \in \mathbb{C} \). 
		\newline
		Since
		\bdm	
			\mathcal{G}_+(\vec{x}) =  \int_0^{\infty} dk \sum_{l=0}^{\infty} \sum_{m = -l}^{l} \bigg[ \frac{1}{k^2 - m \omega + i} + \frac{2i}{|k^2-m \omega - i|^2} \bigg] \: \varphi^*_{klm} (\vec{y}_0) \: \varphi_{klm}(\vec{x}) =
		\edm
		\bdm
			= \mathcal{G}_- (\vec{x}) + 2i g(\vec{x}, \vec{y}_0)
		\edm
		where 
		\bdm
			g(\vec{x}, \vec{y}_0) =  \int_0^{\infty} dk \sum_{l=0}^{\infty} \sum_{m = -l}^{l} \frac{1}{|k^2-m \omega - i|^2} \: \varphi^*_{klm} (\vec{y}_0) \: \varphi_{klm}(\vec{x})
		\edm
		belongs to \( \mathcal{D}(\dot{H}_{\omega}) \), \( \forall \vec{y}_0 \in \mathbb{R}^3 \), we obtain
		\bdm
			\Psi(\vec{x}) = f(\vec{x}) + 2i c \: g(\vec{x}, \vec{y}_0) + c (1 + e^{i \alpha}) \mathcal{G}_-(\vec{x})
		\edm
		and
		\bdm
			\lim_{\vec{x} \rightarrow \vec{y}_0} \Big[ \Psi(\vec{x}) - c (1 + e^{i \alpha}) \mathcal{G}_-(\vec{x}) \Big] = 2i c \| \mathcal{G}_-(\vec{x}) \|^2_{L^2} 
		\edm
		Thus \( \Psi \) can be uniquely decomposed in
		\bdm
			\Psi(\vec{x}) = \Phi(\vec{x}) + \lambda(-i, \alpha) \Phi(\vec{y}_0) \mathcal{G}_-(\vec{x})
		\edm
		with \( \Phi \in \mathcal{D}(\dot{H}_{\omega}) \) and boundary condition
		\bdm
			\lim_{\vec{x} \rightarrow \vec{y}_0} \Big[ \Psi(\vec{x}) - \lambda(-i, \alpha) \Phi(\vec{y}_0) \mathcal{G}_-(\vec{x}) \Big] = \Phi(\vec{y}_0) 
		\edm
		Comparing the two boundary conditions we obtain
		\bdm
			\Phi(\vec{y}_0) = 2i c \| \mathcal{G}_-(\vec{x}) \|^2_{L^2}
		\edm
		\bdm
			c (1 + e^{i \alpha}) = \lambda(-i, \alpha) \Phi(\vec{y}_0)
		\edm
		and then
		\bdm
			c = \frac{ \Phi(\vec{y}_0)}{2i  \| \mathcal{G}_-(\vec{x}) \|^2_{L^2}}
		\edm
		\bdm
			\lambda(-i, \alpha) = \frac{1 + e^{i \alpha}}{2i  \| \mathcal{G}_-(\vec{x}) \|^2_{L^2}}
		\edm
		\begin{flushright} 
			\( \Box \)
		\end{flushright}

	\begin{teo}
		\label{Spe1}
		The spectrum \( \sigma(K_{\alpha, y_0}) \) is purely absolutely continuous and
		\martin
			 \sigma(K_{\alpha, y_0}) = \sigma_{\mathrm{ac}}(K_{\alpha, y_0}) = \sigma(H_{\omega}) = \mathbb{R}
		\end{equation}
	\end{teo}
		
	\emph{Proof:} 	Considering the explicit expression of the resolvent given in Theorem \ref{Res2}, we immediately see that \( \sigma(K_{\alpha, y_0}) = \sigma(H_{\omega}) = \mathbb{R} \): indeed, since \( (K_{\alpha, y_0} - z)^{-1} - (H_{\omega} - z)^{-1} \) is of rank 1 for each \( z \in \mathbb{R} \) and \( \alpha \in [0, 2\pi) \), Weyl's Theorem (see for example Theorem XIII.14 in \cite{Reed1}) implies \( \sigma_{\mathrm{ess}}(K_{\alpha, y_0}) = \sigma_{\mathrm{ess}}(H_{\omega}) \).
		\newline
		In order to prove absence of pure point and singular spectrum, we are going to apply the limiting absorption principle (see Theorem XIII.19 in \cite{Reed1}): to this purpose we need to prove that the following inequality is satisfied for every interval \( [a,b] \subset \mathbb{R} \),
		\bdm
			\sup_{0 < \varepsilon < 1} \int_a^b dx \:\: \Big| \Im\Big[ \Big( \Psi \: , \: \big( K_{\alpha, y_0} - x - i \varepsilon \big)^{-1} \Psi \Big) \Big] \Big|^p < \infty
		\edm
		with \( \Psi \) in a dense subset of \( L^2(\mathbb{R}^3) \) and \( p > 1 \).
		\newline
		Since the operator \( H_{\omega} \) has no singular spectrum, the inequality is easily satisfied if \( \alpha = \pi \). So, let \( \alpha \neq \pi \), from Theorem \ref{Res2} one has
		\bdm
			\Big( \Psi \: , \: \big( K_{\alpha, y_0} - x - i \varepsilon \big)^{-1} \Psi \Big) = \Big( \Psi \: , \: \big( H_{\omega} - x - i \varepsilon \big)^{-1} \Psi \Big) + 
		\edm
		\bdm
			+ \lambda(\alpha, x+i\varepsilon) \: \Big( \mathcal{G}_{x-i\varepsilon} \: , \: \Psi \Big) \: \Big( \Psi \: , \: \mathcal{G}_{x+i\varepsilon} \Big)
		\edm
		and again the inequality holds for the first term. It is very easy to see that the second term is a bounded function of \( x \) if \( \varepsilon > 0 \), so that we have only to control the limit when \( \varepsilon \rightarrow 0 \). Since the singular spectrum of \( H_{\omega} \) is empty, we can choose the dense subset of \( L^2(\mathbb{R}^3) \) given by functions of the form \( (H_{\omega} - x) \varphi \) where \( \varphi \in \mathcal{D}(H_{\omega}) \):
		\bdm
			\Big( \mathcal{G}_{x-i\varepsilon} \: , \: \Psi \Big) \: \Big( \Psi \: , \: \mathcal{G}_{x+i\varepsilon} \Big) = \Big[ \big( H_{\omega} - x - i \varepsilon \big)^{-1} \big( H_{\omega} - x \big) \varphi \Big] (\vec{y}_0) \: \cdot
		\edm
		\bdm
			\cdot \: \Big[ \big( H_{\omega} - x - i \varepsilon \big)^{-1} \big( H_{\omega} - x \big) \varphi^* \Big] (\vec{y}_0) \: \underset{\varepsilon \rightarrow 0}{\longrightarrow} \: \big| \varphi(\vec{y}_0) \big|^2 < \infty
		\edm
		since functions in \( \mathcal{D}(H_{\omega}) \) are continuous and because
		\bdm
			\Big[ \big( H_{\omega} - x - i \varepsilon \big)^{-1} \big( H_{\omega} - x \big) \varphi \Big] (\vec{y}_0) = \varphi(\vec{y}_0) + i \varepsilon \Big[ \big( H_{\omega} - x - i \varepsilon \big)^{-1} \varphi \Big] (\vec{y}_0)
		\edm
		and
		\bdm
			\lim_{\varepsilon \rightarrow 0} \Big| \varepsilon \Big[ \big( H_{\omega} - x - i \varepsilon \big)^{-1} \varphi \Big] (\vec{y}_0) \Big| \leq \lim_{\varepsilon \rightarrow 0} \varepsilon \: \big\| \mathcal{G}_{x-i \varepsilon} \big\| \: \| \varphi \| = 0 
		\edm
		Indeed from Proposition \ref{Gre1} we can easily extract the following upper bound for \( \big\| \mathcal{G}_{x-i \varepsilon} \big\| \),
		\bdm
			\big\| \mathcal{G}_{x-i \varepsilon} \big\| \leq \frac{C}{\sqrt{\varepsilon}}
		\edm
		Finally from equation (\ref{lambda1}) it follows that
		\bdm
			\big| \lambda(\alpha, x + i \varepsilon) \big| \underset{\varepsilon \rightarrow 0}{\longrightarrow} 0 
		\edm
		Since the previous argument applies for each interval \( [a,b] \subset \mathbb{R} \), the proof is completed.
		\begin{flushright} 
			\( \Box \)
		\end{flushright}	
	
\subsection{Asymptotic Limit of Rapid Rotation}	

Let \( U_{\mathrm{rot}}(t-s) \) the unitary group generated by \( K_{\alpha, y_0} \) for some \( \alpha \in [0,2\pi) \), according to \cite{Enss1},
	\bdm
		U_{\mathrm{inert}}(t,s) = R(t) \: U_{\mathrm{rot}}(t-s) \: R^{\dagger}(s)
	\edm
In the following, we shall prove that 
	\bdm
		\mathrm{s-}\lim_{\omega \rightarrow \infty} U_{\mathrm{inert}}(t,s) = e^{-iH_{\gamma,C} (t-s)}
	\edm
where \( H_{\gamma,C} \) is an appropriate self-adjoint extension of \( H_C \), a singular perturbation of the Laplacian supported over a circle of radius \( y_0 \) in the \( x,y-\)plane: let \( C \) the curve \( \vec{y}(\varphi) = (y_0, \frac{\pi}{2}, \varphi) \), \( \varphi \in [0, 2\pi] \), and \( \dot{H}_{C} \) the closure of the operator
	\bdm
		H_{C} = H_0
	\edm
	\bdm
		\mathcal{D}(H_{C}) = C^{\infty}_0(\mathbb{R}^3 - C)
	\edm
	we first classify the self-adjoint extensions of \( \dot{H}_C \):
 
	\begin{pro}
		The self-adjoint extensions of the operator \( \dot{H}_{C} \), that are invariant under rotations around the \(z-\)axis,  are given by the one-parameter family \( H_{\gamma, C} \), \( \gamma  \in \mathbb{R} \), with domain
		\bdm
			\mathcal{D}(H_{\gamma, C}) = \{ \Psi \in L^2(\mathbb{R}^3) \: | \: \exists \: \xi_{\Psi} \in \mathcal{D}(\Gamma_{\gamma, C}(z)), \Psi - \tilde{G}_{z} \xi_{\Psi} \in H^2(\mathbb{R}^3), 
		\edm
		\martin
			\big( \Psi - \tilde{G}_{z} \xi_{\Psi} \big)\big|_C = \Gamma_{\gamma, C}(z) \xi_{\Psi} \}
		\sileno
		\martin
			\big( H_{\gamma, C} - z \big) \Psi = \big( H_0 - z \big) \big( \Psi - \tilde{G}_{z} \xi_{\Psi} \big)
		\sileno
		where \( z \in \mathbb{C} \), \( \Im(z) > 0 \), 
		\martin
			\mathcal{D}(\Gamma_{\gamma, C}(z)) = \{ \xi \in L^2([0, 2 \pi]) \: | \: \Gamma_{\gamma, C}(z)_m \xi_m \in l^2 \}
		\sileno
		\bdm
			\xi_m \equiv \frac{1}{\sqrt{2 \pi}} \int_0^{2 \pi} d \phi \: \xi(\phi) e^{-im\phi}
		\edm
		\martin
			\big( \Gamma_{\gamma, C}(z) \xi \big)(\phi) = \gamma \xi(\phi) - \int_0^{2 \pi} d\phi^{\prime} \frac{e^{i \sqrt{z}|\vec{y}(\phi) - \vec{y}(\phi^{\prime})|}}{4 \pi |\vec{y}(\phi) - \vec{y}(\phi^{\prime})|} \xi(\phi^{\prime})
		\sileno
		\martin
			\Gamma_{\gamma, C}(z)_m = \gamma - 2 \pi \int_0^{\infty} dk \sum_{l=|m|}^{\infty} \frac{1}{k^2- z} \big| \varphi_{klm}(\vec{y}_0) \big|^2
		\sileno
		and
		\bdm
			\big( \tilde{G}_{z} \xi \big)(\vec{x}) \equiv \int_0^{2 \pi}  d \phi \: \frac{e^{i \sqrt{z} |\vec{x} - \vec{y}(\phi)|}}{4 \pi |\vec{x} - \vec{y}(\phi)|} \xi(\phi)
		\edm
	\end{pro}

	\emph{Proof:}	See \cite{Teta1, Teta2}. The formula for \( \Gamma_{\alpha, C}(\lambda)_m \) is obtained expressing the free resolvent in terms of spherical waves.
		\begin{flushright} 
			\( \Box \)
		\end{flushright}

	\begin{pro}
		\label{ReH1}
		For every \( \Psi  \in L^2(\mathbb{R}^3) \), \( z \in \varrho(H_{\gamma, C}) \), \( \Im (z) > 0 \) and \( \vec{y}_0 = (0,y_0,0) \),
		\bdm
			\big(H_{\gamma, C} - z \big)^{-1} \Psi (\vec{x}) = \big(H_0 - z \big)^{-1} \Psi (\vec{x}) \: +
		\edm
		\bdm
			+ \sum_{m=-\infty}^{+\infty} \frac{2 \pi}{\Gamma_{\gamma, C}(z)_m} \: G_z^m(\vec{x}, \vec{y}_0) \: \Big( {G_z^m}^*(\vec{x}^{\prime}, \vec{y}_0) \:, \Psi (\vec{x}^{\prime}) \Big)_{L^2(\mathbb{R}^3)}
		\edm
		where
		\bdm
			G^{m}_z (\vec{x}, \vec{y}_0) \equiv \int_0^{\infty} dk \sum_{l=|m|}^{\infty} \frac{1}{k^2 - z} \: \varphi^*_{klm}(\vec{y}_0) \: \varphi_{klm}(\vec{x}) 
		\edm
	\end{pro}

	\emph{Proof:}	The expression for the resolvent of \( H_{\gamma, C} \) for a generic curve \( C \) is given in \cite{Teta1, Teta2}:
		\bdm
			\big(H_{\gamma, C} - z \big)^{-1} \Psi (\vec{x}) =  \big(H_0 - z \big)^{-1} \Psi (\vec{x}) + \tilde{G}_{z} \bigg[ {\Gamma^{-1}_{\gamma, C}(z)} \Big( \big( H_0-z \big)^{-1} \Psi \Big) \Big|_C \bigg]
		\edm
	Since \( \Gamma_{\gamma, C}(z) \) is diagonal in the basis \( e_m (\phi) = \frac{1}{\sqrt 2\pi} e^{im \phi} \) of \( L^2([0, 2 \pi], d\phi) \),
		\bdm
			\big( {\Gamma^{-1}_{\gamma, C}(z)} \xi \big) (\phi) = \sum_{m=-\infty}^{\infty} \frac{1}{\Gamma_{\gamma, C}(z)_m} \: \xi_m \: e_m(\phi)
		\edm
		and therefore	
		\bdm
			{\Gamma^{-1}_{\gamma, C}(z)} \Big( \big( H_0-z )^{-1} \Psi \Big) \Big|_C = \sum_{m=-\infty}^{\infty} \frac{\Big[ \Big( \big( H_0-z )^{-1} \Psi \Big) \Big|_C \Big]_m}{\Gamma_{\gamma, C}(z)_m} \:  e_m(\phi)
		\edm
		where
		\bdm
			\Big[ \Big( \big( H_0-z )^{-1} \Psi \Big) \Big|_C \Big]_m = \frac{1}{\sqrt{2 \pi}} \int_0^{2 \pi} d \phi \: e^{-im\phi} \int_{\mathbb{R}^3} d^3 \vec{x}^{\prime} \int_0^{\infty} dk \sum_{l=0}^{\infty} \sum_{m^{\prime} = -l}^{l} \frac{1}{k^2-z} \cdot
		\edm
		\bdm
			\cdot \: \varphi^*_{klm^{\prime}}(\vec{x}^{\prime}) \: \varphi_{klm^{\prime}}(\vec{y}(\phi)) \: \Psi(\vec{x}^{\prime}) = \sqrt{2 \pi} \: e^{im \frac{\pi}{2}} \int_{\mathbb{R}^3} d^3 \vec{x}^{\prime} \: G^{m}_z (\vec{x}^{\prime}, \vec{y}_0) \: \Psi(\vec{x}^{\prime})
		\edm
		Finally
		\bdm
			(\tilde{G}_{z} \: e_m)(\vec{x}) = \int_0^{2 \pi} \frac{d \phi}{\sqrt{2 \pi}} \int_0^{\infty} dk \sum_{l=0}^{\infty} \sum_{m^{\prime} = -l}^{l} \frac{1}{k^2-z} \: \varphi^*_{klm^{\prime}}(\vec{y}(\phi)) \: \varphi_{klm^{\prime}}(\vec{x}) \: e^{im\phi} = 
		\edm
		\bdm
			= \sqrt{2 \pi}  \: e^{-im \frac{\pi}{2}} \: G^{m}_z (\vec{x}, \vec{y}_0)
		\edm
		\begin{flushright} 
			\( \Box \)
		\end{flushright}

	\begin{cor}
		\label{ReH2}
		If  \( \Psi(\vec{x}) \in L^2(\mathbb{R}^3) \), \( \Psi(\vec{x}) = \chi(r) Y_{l_0}^{m_0}(\theta, \phi) \) and \( z\in \varrho(H_{\gamma, C}) \), \linebreak \( \Im(z) > 0 \),
		\bdm
			\Big( \big(H_{\gamma, C} - z \big)^{-1} \Psi \Big) (\vec{x}) = \int_0^{\infty} dr^{\prime} {r^{\prime}}^2 \:  g_z^{l_0}(r,r^{\prime}) \chi(r^{\prime}) \:\:  Y_{l_0}^{m_0} (\theta, \phi) \: + 
		\edm
		\bdm
			+ \: \frac{2 \pi \: Y_{l_0}^{m_0}(\pi/2, 0)}{\Gamma_{\gamma, C}(z)_{m_0}} \: G^{m_0}_z(\vec{x}, \vec{y}_0) \int_0^{\infty}  dr^{\prime} {r^{\prime}}^2 \: g_z^{l_0} (y_0, r^{\prime}) \chi(r^{\prime})
		\edm
		where
		\bdm
			g_z^{l_0} (r, r^{\prime}) \equiv \frac{2}{\pi} \int_0^{\infty} dk \: \frac{k^2}{k^2 - z} \: j_{l_0}(kr) j_{l_0}(kr^{\prime}) = \big( H_0 - z \big)^{-1} \big|_{\mathcal{H}_{l_0}^{m_0}} (r,r^{\prime})
		\edm
		and \( \mathcal{H}_{l_0}^{m_0} \) is the subspace of \( L^2(\mathbb{R}^3) \) spanned by \( \chi(r) Y_{l_0}^{m_0}(\theta, \phi) \).
	\end{cor}

	\emph{Proof:}	The result follows from a straightforward calculation: indeed, if \( \Psi(\vec{x}) = \chi(r) Y_{l_0}^{m_0}(\theta, \phi) \),
		\bdm
			\Big( {G_z^m}^*(\vec{x}^{\prime}, \vec{y}_0) \:, \Psi(\vec{x}^{\prime}) \Big) = \delta_{m,m_0} \: Y_{l_0}^{m_0} (\pi/2, 0)  \int_0^{\infty}  dr^{\prime} {r^{\prime}}^2 \: g_z^{l_0} (y_0, r^{\prime}) \chi(r^{\prime})
		\edm
		and
		\bdm
			\Big( \big(H_0 - z \big)^{-1} \Psi \Big) (\vec{x}) = \int_0^{\infty} dr^{\prime} {r^{\prime}}^2 \:  g_z^{l_0}(r,r^{\prime}) \chi(r^{\prime}) \:\:  Y_{l_0}^{m_0} (\theta, \phi)
		\edm
		\begin{flushright} 
			\( \Box \)
		\end{flushright}	

Now we can state the main result:

	\begin{teo}
		\label{Asy1}
		For every \( t,s \in \mathbb{R} \),
		\bdm
			\mathrm{s-}\lim_{\omega \rightarrow \infty} U_{\mathrm{inert}}(t,s) = e^{-iH_{\gamma, C} (t-s)}
		\edm
		where \( \gamma (\alpha, y_0) \in \mathbb{R} \) and
		\bdm
			\gamma(\alpha, y_0) = 2 \pi \int_0^{\infty} dk \sum_{l=0}^{\infty} \bigg[ \frac{2i}{(1+e^{i\alpha})|k^2+i|^2} + \frac{1}{k^2+i} \bigg] \big| \varphi_{kl0}(\vec{y}_0) \big|^2
		\edm
	\end{teo}

	\emph{Proof:}	First we observe that (see Lemma \ref{Con1} below)
		\bdm
			\mathrm{s-}\lim_{\omega \rightarrow \infty} \int_{-\infty}^0 dt \: e^{-izt} \: U^{*}_{\mathrm{inert}}(t,0) = -i \big(H_{\gamma,C} - z \big)^{-1} = \int_{-\infty}^0 dt \: e^{-izt} \:  e^{iH_{\gamma, C} t}
		\edm
		and, since the previous equality holds for every \( z \in \mathbb{C} \), \( \Im(z) > 0 \), we obtain
		\bdm
			\mathrm{s-}\lim_{\omega \rightarrow \infty} U^{*}_{\mathrm{inert}}(t,0) = e^{iH_{\gamma, C} t}
		\edm
		and therefore
		\bdm
			\mathrm{s-}\lim_{\omega \rightarrow \infty} U_{\mathrm{inert}}(t,0) = e^{-iH_{\gamma, C} t}
		\edm
		The result then follows from the property of the 2-parameters unitary group \( U_{\mathrm{inert}}(t,s) \):
		\bdm
			\mathrm{s-}\lim_{\omega \rightarrow \infty} U_{\mathrm{inert}}(t,s) = \mathrm{s-}\lim_{\omega \rightarrow \infty} \bigg[ U_{\mathrm{inert}}(t,0) \: U_{\mathrm{inert}}^*(s,0) \bigg] = e^{-iH_{\gamma, C} (t-s)}
		\edm
		\begin{flushright} 
			\( \Box \)
		\end{flushright}

		The explicit expression of the parameter \( \gamma(\alpha, y_0) \) is proved in the following Lemma \ref{Con1}.

	\begin{lem}
		\label{Con1}
		For every \( z \in \mathbb{C} \), \( \Im(z) > 0 \),
		\bdm
			\mathrm{s-}\lim_{\omega \rightarrow \infty} \int_{-\infty}^0 dt \: e^{-izt} \: U^{*}_{\mathrm{inert}}(t,0) = -i \big(H_{\gamma,C} - z \big)^{-1}
		\edm
	\end{lem}

	\emph{Proof:}	We shall verify the equality on the dense subset of \( L^2(\mathbb{R}^3) \) given by functions of the form \( \Psi(\vec{x}) = \chi(r) Y_{l_0}^{m_0}(\theta, \phi) \), with \( l_0 = 0, \ldots \infty \) and \( m_0 = -l_0, \ldots, l_0 \),
		\bdm
			U^{*}_{\mathrm{inert}}(t,0) \Psi(\vec{x}) =  e^{i K_{\alpha,y_0} t} R^{*}(t) \Psi(\vec{x}) = e^{i (K_{\alpha,y_0}+ m_0 \omega) t} \Psi(\vec{x})
		\edm
		Therefore
		\bdm
	 		\int_{-\infty}^0 dt \: e^{-izt} \: U^{*}_{\mathrm{inert}}(t,0) \Psi(\vec{x}) =  \int_{-\infty}^0 dt \: e^{-izt} \: e^{i (K_{\alpha,y_0}+ m_0 \omega) t} \Psi(\vec{x}) = 
		\edm
		\bdm
			\int_{-\infty}^0 dt \: e^{-i(z-m_0 \omega)t} \: e^{i K_{\alpha,y_0} t} \Psi(\vec{x}) = -i \big( K_{\alpha,y_0} +m_0 \omega - z \big)^{-1} \Psi(\vec{x}) 
		\edm
		Hence we have now to prove that
		\bdm
			\lim_{\omega \rightarrow \infty} \big( K_{\alpha,y_0}+m_0 \omega - z \big)^{-1} \Psi(\vec{x}) =  \big(H_{\gamma,C} - z \big)^{-1} \Psi(\vec{x})
		\edm
		First of all we observe that, for each \( z \in \mathbb{C} \), \( \Im(z) > 0 \), \( m_0 \in \mathbb{Z} \) and \( \vec{y}_0 = (0,y_0,0) \),
		\bdm
			\lim_{\omega \rightarrow \infty} \mathcal{G}_{z-m_0 \omega}(\vec{x}, \vec{y}_0) = G^{m_0}_z (\vec{x}, \vec{y}_0)
		\edm	
		in the norm topology of \( L^2(\mathbb{R}^3) \): indeed, since
		\bdm
			\mathcal{G}_{z-m_0 \omega}(\vec{x}, \vec{y}_0) = G^{m_0}_z (\vec{x}, \vec{y}_0) + R_z^{m_0} (\vec{x}, \vec{y}_0)
		\edm
		with
		\bdm
			R_z^{m_0} (\vec{x}, \vec{y}_0) = \int_0^{\infty} dk \sum_{l=0}^{\infty} \underset{m \neq m_0}{\sum_{m=-l}^{l}} \frac{1}{k^2 -(m-m_0) \omega - z} \: \varphi^*_{klm}(\vec{y}_0) \: \varphi_{klm}(\vec{x})
		\edm
		it is sufficient to prove that
		\bdm
			\lim_{\omega \rightarrow \infty} \big\| R_z^{m_0} (\vec{x},\vec{y}_0) \big\|_{L^2(\mathbb{R}^3)} = 0
		\edm
		but
		\bdm
			\big\| R_z^{m_0} (\vec{x},\vec{y}_0) \big\|^2_{L^2(\mathbb{R}^3)} = \int_0^{\infty} dk \sum_{l=0}^{\infty} \underset{m \neq m_0}{\sum_{m=-l}^{l}} \frac{1}{|k^2 -(m-m_0) \omega - z|^2} |\varphi_{klm}(\vec{y}_0)|^2
		\edm
		and the right hand side is bounded for each \( \omega \in \mathbb{R} \) (see Proposition \ref{Gre1}), so that we can exchange the limit with the integration
		\bdm
			\lim_{\omega \rightarrow \infty} \int_0^{\infty} dk \sum_{l=0}^{\infty} \underset{m \neq m_0}{\sum_{m=-l}^{l}} \frac{1}{|k^2 -(m-m_0) \omega - z|^2} |\varphi_{klm}(\vec{y}_0)|^2 = 
		\edm
		\bdm
			= \int_0^{\infty} dk \sum_{l=0}^{\infty} \underset{m \neq m_0}{\sum_{m=-l}^{l}} |\varphi_{klm}(\vec{y}_0)|^2 \lim_{\omega \rightarrow \infty} \frac{1}{|k^2 -(m-m_0) \omega - z|^2} = 0
		\edm
		Now, since (see Theorem \ref{Res2})
		\bdm
			\Big[ \big( K_{\alpha,y_0}+m_0 \omega - z \big)^{-1} \Psi \Big] (\vec{x}) = \Big( \mathcal{G}^*_{z-m_0 \omega}(\vec{x}, \vec{x}^{\prime}),  \Psi(\vec{x}^{\prime}) \Big)_{L^2(\mathbb{R}^3)} + 
		\edm
		\bdm
			+ \lambda(z-m_0 \omega, \alpha) \Big( \mathcal{G}_{\bar{z}-m_0 \omega}(\vec{x}^{\prime}, \vec{y}_0), \Psi(\vec{x}^{\prime}) \Big)_{L^2(\mathbb{R}^3)} \mathcal{G}_{z-m_0 \omega}(\vec{x},\vec{y}_0) 
		\edm
		and
		\bdm
			\lim_{\omega \rightarrow \infty} \Big( \mathcal{G}^*_{z-m_0 \omega}(\vec{x}, \vec{x}^{\prime}),  \Psi(\vec{x}^{\prime}) \Big)_{L^2(\mathbb{R}^3)} = e^{i m_0 \frac{\pi}{2}} \Big( {G^{m_0}_z}^* (\vec{x}, \vec{x}^{\prime}) ,  \Psi(\vec{x}^{\prime}) \Big)_{L^2(\mathbb{R}^3)} =
		\edm
                \bdm
			= \int_0^{\infty}  dr^{\prime} {r^{\prime}}^2 g_z^{l_0}(y_0,r^{\prime}) \: \chi(r^{\prime}) \:\: Y_{l_0}^{m_0} (\pi/2, \pi/2)
		\edm
		\bdm
			\lim_{\omega \rightarrow \infty} \Big( \mathcal{G}_{\bar{z}-m_0 \omega}(\vec{x}^{\prime}, \vec{y}_0),  \Psi(\vec{x}^{\prime}) \Big)_{L^2(\mathbb{R}^3)} = e^{-i m_0 \frac{\pi}{2}} \Big(G^{m_0}_{\bar{z}} (\vec{x}^{\prime}, \vec{y}_0),  \Psi(\vec{x}^{\prime}) \Big)_{L^2(\mathbb{R}^3)} 
		\edm
		\bdm
			\lim_{\omega \rightarrow \infty}  \mathcal{G}_{z-m_0 \omega}(\vec{x},\vec{y}_0) = e^{i m_0 \frac{\pi}{2}} G^{m_0}_z (\vec{x}, \vec{y}_0)
		\edm
		we obtain
		\bdm	
			\lim_{\omega \rightarrow \infty} \big( K_{\alpha,y_0}+m_0 \omega - z \big)^{-1} \Psi(\vec{x}) = \int_0^{\infty} dr^{\prime} {r^{\prime}}^2 \:  g_z^{l_0}(r,r^{\prime}) \chi(r^{\prime}) \:\:  Y_{l_0}^{m_0} (\theta, \phi) \: +
		\edm
		\bdm
			+ \: \beta(z, \alpha) \: G^{m_0}_z(\vec{x}, \vec{y}_0) \int_0^{\infty}  dr^{\prime} {r^{\prime}}^2 \: g_z^{l_0} (y_0, r^{\prime}) \chi(r^{\prime}) = \big(H_{\gamma,C} - z \big)^{-1} \Psi(\vec{x})
		\edm
		with\footnote{Actually \( \lambda \) is a function separately of \( z - m_0 \omega \) and \( \omega \), since the Green's function \( \mathcal{G}_-(\vec{x}) \) depends on \( \omega \).}
		\bdm
			\beta(z, \alpha) = \lim_{\omega \rightarrow \infty}  \lambda(z-m_0 \omega, \alpha)
		\edm
		and
		\bdm
			\frac{\Gamma_{\gamma, C}(z)_{m_0}}{2 \pi} = \frac{1}{\beta(z, \alpha)}	
		\edm
		It remains to find the explicit expression of \( \gamma(\alpha, y_0) \): using the relation (see Theorem \ref{Res2})
		\bdm
			\frac{1}{\lambda(z-m_0 \omega, \alpha)} = \frac{1}{\lambda(-i, \alpha)} - (z-m_0 \omega + i) \Big( \mathcal{G}_{-m_0 \omega +\bar{z}} (\vec{x}) , \: \mathcal{G}_{-} (\vec{x}) \Big)
		\edm
		we obtain
		\bdm
			\frac{1}{\beta(z, \alpha)} =  \lim_{\omega \rightarrow \infty} \bigg[ \frac{1}{\lambda(-i, \alpha)} - (z-m_0 \omega + i) \Big( \mathcal{G}_{-m_0 \omega +\bar{z}} (\vec{x}) , \: \mathcal{G}_{-} (\vec{x}) \Big) \bigg] =
		\edm
		\bdm
			= \frac{2i}{1+e^{i \alpha}} \int_0^{\infty} dk \sum_{l=0}^{\infty} \frac{1}{|k^2+i|^2} \big| \varphi_{kl0}(\vec{y}_0) \big|^2 + \int_0^{\infty} dk \sum_{l=0}^{\infty} \frac{1}{k^2+i} \big| \varphi_{kl0}(\vec{y}_0) \big|^2 +
		\edm
		\bdm
			- \int_0^{\infty} dk \sum_{l=|m_0|}^{\infty} \frac{1}{k^2-z} \big| \varphi_{klm_0}(\vec{y}_0) \big|^2
		\edm
		and hence the result. We want to stress that, as it was expected, \( \gamma \in \mathbb{R} \):
		\bdm	
			\Im \bigg\{ \frac{2i}{(1+e^{i\alpha})|k^2+i|^2} + \frac{1}{k^2+i} \bigg\} = \frac{1}{|k^2+i|^2} \bigg\{ \Im \bigg[\frac{2i}{1+e^{i\alpha}} \bigg] - 1 \bigg\} = 
		\edm
		\bdm
			= \frac{1}{|k^2+i|^2} \bigg\{ \frac{\Im \big[2i+2ie^{-i\alpha} \big]}{2+2\cos\alpha} - 1 \bigg\} = 0
		\edm
		\begin{flushright} 
			\( \Box \)
		\end{flushright}

\section{The Rotating Point Interaction in 2D}

\subsection{The Hamiltonian}

The system we shall study is defined by the formal time-dependent Hamiltonian
	\martin
		H(t) = H_0 + a \: \delta^{(2)}(\vec{x} - \vec{y}(t))
	\end{equation}
where \( \vec{y}(t) = \mathcal{R}(t) \vec{y}_0 \).
\newline
The formal generator of time evolution in the uniformly rotating frame (with angular velocity \( \omega \)) is given by
	\bdm
		K = H_0 - \omega J + a \: \delta^{(2)}(\vec{x} - \vec{y}_0)
	\edm
Therefore the Hamiltonian of the system is a self-adjoint extension of the operator
	\bdm
		K_{y_0} = H_{\omega}
	\edm
	\bdm
		\mathcal{D}(K_{y_0}) = C^{\infty}_0 (\mathbb{R}^2 - \{\vec{y}_0\})
	\edm
According to the discussion of Section 2, the Hamiltonian is given by the self-adjoint operator
	\martin
		\mathcal{D}(K_{\alpha,y_0}) = \{ f + c \mathcal{G}_+ + c e^{i\alpha} \mathcal{G}_- | g \in \mathcal{D}(\dot{K}_{y_0}), c \in \mathbb{C} \}
	\end{equation}
	\martin
		K_{\alpha,y_0} ( f + c \mathcal{G}_+ + c e^{i\alpha} \mathcal{G}_- ) = \dot{K}_{y_0} g + i c \mathcal{G}_+ - i c e^{i\alpha} \mathcal{G}_-
	\end{equation}
with \( \alpha \in [0, 2\pi) \) and where
	\bdm
		\mathcal{G}_{\pm}(\vec{x}) = \mathcal{G}_{\pm i} (\vec{x}, \vec{y}_0)  
	\edm
	\martin
		\label{Res3}
		\mathcal{G}_z(\vec{x},\vec{y}_0) = \int_0^{\infty} dk \sum_{n=-\infty}^{\infty} \frac{1}{k^2 - n \omega - z} \: \varphi^*_{kn} (\vec{y}_0) \: \varphi_{kn}(\vec{x}) 
	\end{equation}
for \( \vec{x} \in \mathbb{R}^2 - \{\vec{y}_0\} \).
\newline
As in the 3D case, the self-adjoint extension \( K_{\pi,y_0} \) corresponds to the ``free'' Hamiltonian \( \dot{H}_{\omega} \) and 
	\begin{teo}
		\label{Res4}
		The resolvent of \( K_{\alpha, y_0} \) has integral kernel given by
		\martin	
			(K_{\alpha,y_0} - z)^{-1} (\vec{x}, \vec{x}') = \mathcal{G}_z(\vec{x}, \vec{x}') + \lambda(z, \alpha) \mathcal{G}^*_{\bar{z}}(\vec{x}', \vec{y}_0) \mathcal{G}_z(\vec{x},\vec{y}_0)
		\end{equation}
with \( z \in \varrho(K_{\alpha,y_0}) \), \( \vec{x}, \vec{x}' \in \mathbb{R}^2 \), \( \vec{x} \neq \vec{x}' \), \( \vec{x}, \vec{x}' \neq \vec{y}_0 \) and
		\martin
			\frac{1}{\lambda(z, \alpha)} = \frac{1}{\lambda(-i, \alpha)} - (z+i) \big( \mathcal{G}_{\bar{z}} (\vec{x}) , \mathcal{G}_- (\vec{x}) \big)
		\end{equation}
		\martin
			\lambda(-i, \alpha) = \frac{1 + e^{i\alpha}}{2i \| \mathcal{G}_- (\vec{x}) \|^2}
		\end{equation}
	\end{teo}

	\emph{Proof:} 	See the Proof of Theorem \ref{Res2} and Proposition \ref{Gre2}.
		\begin{flushright} 
			\( \Box \)
		\end{flushright}
	
	\begin{teo}
		\label{Dom2}
		The domain \( \mathcal{D}(K_{\alpha, y_0}) \), \( \alpha \in [0,2\pi) \), consists of all elements \linebreak \( \Psi \in \mathbb{R}^3 \) which can be decomposed in the following way
		\bdm
			\Psi (\vec{x}) = \Phi_z(\vec{x}) + \lambda(z, \alpha) \Phi_z(\vec{y}_0) \mathcal{G}_z(\vec{x}, \vec{y}_0)
		\edm
		for \( \vec{x} \neq \vec{y}_0 \), \( \Phi_z \in \mathcal{D}(\dot{H}_{\omega}) \) and \( z \in \varrho(K_{\alpha, y_0}) \). The previous decomposition is unique and on every \( \Psi \) of this form we obtain
		\bdm
			(K_{\alpha,y_0} - z) \Psi = (H_{\omega} -z) \Phi_z
		\edm
	\end{teo}

	\emph{Proof:} 	See the Proof of Theorem \ref{Dom1}.
		\begin{flushright} 
			\( \Box \)
		\end{flushright}

	\begin{teo}
		\label{Spe2}
		The spectrum \( \sigma(K_{\alpha, y_0}) \) is purely absolutely continuous and
		\martin
			 \sigma(K_{\alpha, y_0}) = \sigma_{\mathrm{ac}}(K_{\alpha, y_0}) = \sigma(H_{\omega}) = \mathbb{R}
		\end{equation}
	\end{teo}
		
	\emph{Proof:} 	See the Proof of Theorem \ref{Spe1}, Theorem \ref{Res2} and Proposition \ref{Gre2}.
		\begin{flushright} 
			\( \Box \)
		\end{flushright}	
	
\subsection{Asymptotic Limit of Rapid Rotation}	

As in the 3D case, we shall prove that 
	\bdm
		\mathrm{s-}\lim_{\omega \rightarrow \infty} U_{\mathrm{inert}}(t,s) = e^{-iH_{\gamma,C} (t-s)}
	\edm
where \( H_{\gamma,C} \) is an appropriate self adjoint extension of \( H_C \), a singular perturbation of the Laplacian supported over a circle of radius \( y_0 \): let \( C \) the curve \( \vec{y}(\theta) = (y_0 , \theta) \), \( \theta \in [0, 2\pi] \), and \( \dot{H}_{C} \) the closure of the operator
	\bdm
		H_{C} = H_0
	\edm
	\bdm
		\mathcal{D}(H_{C}) = C^{\infty}_0(\mathbb{R}^2 - C)
	\edm
 
	\begin{pro}
		The self-adjoint extensions of the operator \( \dot{H}_{C} \), that are invariant under rotations around the \(z-\)axis, are given by the one-parameter family of operators \( H_{\gamma, C} \), \( \gamma  \in \mathbb{R} \), with domain
		\bdm
			\mathcal{D}(H_{\gamma, C}) = \{ \Psi \in L^2(\mathbb{R}^2) \: | \: \exists \: \xi_{\Psi} \in \mathcal{D}(\Gamma_{\gamma, C}(z)), \Psi - \tilde{G}_{z} \xi_{\Psi} \in H^2(\mathbb{R}^2), 
		\edm
		\martin
			\big( \Psi - \tilde{G}_{z} \xi_{\Psi} \big)\big|_C = \Gamma_{\gamma, C}(z) \xi_{\Psi} \}
		\sileno
		\martin
			\big( H_{\gamma, C} - z \big) \Psi = \big( H_0 - z \big) \big( \Psi - \tilde{G}_{z} \xi_{\Psi} \big)
		\sileno
		where \( z \in \mathbb{C} \), \( \Im(z) > 0 \), 
		\martin
			\mathcal{D}(\Gamma_{\gamma, C}(z)) = \{ \xi \in L^2([0, 2 \pi]) \: | \: \Gamma_{\gamma, C}(z)_n \xi_n \in l^2 \}
		\sileno
		\bdm
			\xi_n \equiv \frac{1}{\sqrt{2 \pi}} \int_0^{2 \pi} d \theta \: \xi(\theta) e^{-in\theta} = \Big( e_{n} \: , \: \xi_{\Psi} \Big)_{L^2([0, 2\pi] , d\theta)}
		\edm
		\martin
			\big( \Gamma_{\gamma, C}(z) \xi \big)(\theta) \equiv \frac{\xi(\theta)}{\gamma} - \int_0^{2 \pi} d\theta^{\prime} \frac{e^{i \sqrt{z}|\vec{y}(\theta) - \vec{y}(\theta^{\prime})|}}{4 \pi |\vec{y}(\theta) - \vec{y}(\theta^{\prime})|} \xi(\theta^{\prime})
		\sileno
		\martin
			\Gamma_{\gamma, C}(z)_n = \frac{1}{\gamma} - 2 \pi \int_0^{\infty} dk \frac{1}{k^2- z} \big| \varphi_{kn}(\vec{y}_0) \big|^2
		\sileno
		and
		\bdm
			\big( \tilde{G}_{z} \xi \big)(\vec{x}) \equiv \int_0^{2 \pi}  d \theta  \frac{e^{i \sqrt{z} |\vec{x} - \vec{y}(\theta)|}}{4 \pi |\vec{x} - \vec{y}(\theta)|} \xi(\theta)
		\edm
	\end{pro}

	\emph{Proof:}	Singular perturbations of the Laplacian supported on a curve in \( \mathbb{R}^2 \) are analogous to singular perturbations supported on a surface in \( \mathbb{R}^3 \): indeed the quadratic form
		\bdm
			\mathcal{F}(\Psi, \Psi) \equiv \int_{\mathbb{R}^2} d^2\vec{x} \: \big| \nabla \Psi \big|^2 - \int_{C} d \theta \: \gamma(\theta) \big| \Psi(\vec{y}(\theta)) \big|^2
		\edm
		is easily seen to be a closed semibounded quadratic form (see for example \cite{Teta1, Teta2} and the discussion of Section 5) on 
		\bdm	
			\mathcal{D}(F) = \big\{ \Psi \in L^2(\mathbb{R}^2) \: | \: \exists \: \xi_{\Psi} \in L^2(C), \Psi - \tilde{G}_{z} \xi_{\Psi} \in H^1(\mathbb{R}^2) \big\}
		\edm 
		and it can be proved that it is associated to the self-adjoint operator \( H_{\gamma, C} \).
		\begin{flushright} 
			\( \Box \)
		\end{flushright}

	\begin{pro}
		\label{ReH3}
		If  \( \Psi(\vec{x}) \in L^2(\mathbb{R}^2) \), \( \Psi(\vec{x}) = \chi(r) e_{n_0}(\theta) \) and \( z\in \varrho(H_{\gamma, C}) \), \( \Im(z) > 0 \),
		\bdm
			\Big( \big(H_{\gamma, C} - z \big)^{-1} \Psi \Big) (\vec{x}) = \int_0^{\infty} dr^{\prime} r^{\prime} \:  g_z^{n_0}(r,r^{\prime}) \chi(r^{\prime}) \: + 
		\edm
		\bdm
			+ \: \frac{2 \pi}{\Gamma_{\gamma, C}(z)_{n_0}} \: G^{n_0}_z(\vec{x}, \vec{y}_0) \int_0^{\infty}  dr^{\prime} r^{\prime} \: g_z^{n_0} (y_0, r^{\prime}) \chi(r^{\prime})
		\edm
		where
		\bdm
			g_z^{n_0} (r, r^{\prime}) \equiv \int_0^{\infty} dk \: \frac{k}{k^2 - z} \: J_{|n_0|}(kr) J_{|n_0|}(kr^{\prime}) = \big( H_0 - z \big)^{-1} \big|_{\mathcal{H}_{n_0}} (r,r^{\prime})
		\edm
		and
		\bdm
			G^{n}_z (\vec{x}, \vec{y}_0) \equiv \int_0^{\infty} dk \frac{1}{k^2 - z} \: \varphi^*_{kn}(\vec{y}_0) \: \varphi_{kn}(\vec{x}) 
		\edm
	\end{pro}

	\emph{Proof:}	See the Proof of Proposition \ref{ReH1} and Corollary \ref{ReH2}.
		\begin{flushright} 
			\( \Box \)
		\end{flushright}

	\begin{teo}
		For every \( t,s \in \mathbb{R} \),
		\bdm
			\mathrm{s-}\lim_{\omega \rightarrow \infty} U_{\mathrm{inert}}(t,s) = e^{-iH_{\gamma, C} (t-s)}
		\edm
		where \( \gamma (\alpha, y_0) \in \mathbb{R} \) and
		\bdm
			\gamma(\alpha, y_0) = \int_0^{\infty} dk \: k \bigg[ \frac{2i}{(1+e^{i\alpha})|k^2+i|^2} + \frac{1}{k^2+i} \bigg] J^2_{0}(ky_0)
		\edm
	\end{teo}

	\emph{Proof:}	See the Proof of Theorem \ref{Asy1} and the following Lemma \ref{Con2}.
		\begin{flushright} 
			\( \Box \)
		\end{flushright}

	\begin{lem}
		\label{Con2}
		For every \( z \in \mathbb{C} \), \( \Im(z) > 0 \),
		\bdm
			\mathrm{s-}\lim_{\omega \rightarrow \infty} \int_{-\infty}^0 dt \: e^{-izt} \: U^{*}_{\mathrm{inert}}(t,0) = -i \big(H_{\gamma,C} - z \big)^{-1}
		\edm
	\end{lem}

	\emph{Proof:}	The first part of the proof is analogous to the Proof of Lemma \ref{Con1} (the only difference is the dense subset of \( L^2(\mathbb{R}^2) \) given by functions of the form \( \Psi(\vec{x}) = \chi(r) e_{n_0}(\theta) \), with \( n_0 \in \mathbb{Z} \)). 
		\newline
		Hence it remains to prove that
		\bdm
			\lim_{\omega \rightarrow \infty} \big( K_{\alpha,y_0}+n_0 \omega - z \big)^{-1} \Psi(\vec{x}) =  \big(H_{\gamma,C} - z \big)^{-1} \Psi(\vec{x})
		\edm
		Now, for each \( z \in \mathbb{C} \), \( \Im(z) > 0 \), \( n_0 \in \mathbb{Z} \) and \( \vec{y}_0 = (0,y_0) \),
		\bdm
			\lim_{\omega \rightarrow \infty} \mathcal{G}_{z-n_0 \omega}(\vec{x}, \vec{y}_0) = G^{n_0}_z (\vec{x}, \vec{y}_0)
		\edm	
		in the norm topology of \( L^2(\mathbb{R}^2) \): since
		\bdm
			\mathcal{G}_{z-n_0 \omega}(\vec{x}, \vec{y}_0) = G^{n_0}_z (\vec{x}, \vec{y}_0) + R_z^{n_0} (\vec{x}, \vec{y}_0)
		\edm
		with
		\bdm
			R_z^{n_0} (\vec{x}, \vec{y}_0) = \int_0^{\infty} dk \underset{n \neq n_0}{\sum_{n=-\infty}^{\infty}} \frac{1}{k^2 -(n-n_0) \omega - z} \: \varphi^*_{kn}(\vec{y}_0) \: \varphi_{kn}(\vec{x})
		\edm
		it is sufficient to prove that
		\bdm
			\lim_{\omega \rightarrow \infty} \big\| R_z^{n_0} (\vec{x},\vec{y}_0) \big\|_{L^2(\mathbb{R}^2)} = 0
		\edm
		But
		\bdm
			\big\| R_z^{n_0} (\vec{x},\vec{y}_0) \big\|^2_{L^2(\mathbb{R}^2)} = \int_0^{\infty} dk \underset{n \neq n_0}{\sum_{n=-\infty}^{\infty}} \frac{1}{|k^2 -(n-n_0) \omega - z|^2} |\varphi_{kn}(\vec{y}_0)|^2
		\edm
		and the right hand side is bounded (see Proposition \ref{Gre2}) for each \( \omega \in \mathbb{R} \), so that exchanging the limit with the integration, we obtain the result.
		\newline
		Now, substituting in the expression of the resolvent (see Theorem \ref{Res4}),
		\bdm
			\Big[ \big( K_{\alpha,y_0}+m_0 \omega - z \big)^{-1} \Psi \Big] (\vec{x}) = \Big( \mathcal{G}^*_{z-m_0 \omega}(\vec{x}, \vec{x}^{\prime}),  \Psi(\vec{x}^{\prime}) \Big)_{L^2(\mathbb{R}^3)} + 
		\edm
		\bdm
			+ \lambda(z-m_0 \omega, \alpha) \Big( \mathcal{G}_{\bar{z}-m_0 \omega}(\vec{x}^{\prime}, \vec{y}_0), \Psi(\vec{x}^{\prime}) \Big)_{L^2(\mathbb{R}^3)} \mathcal{G}_{z-m_0 \omega}(\vec{x},\vec{y}_0) 
		\edm
		the result follows from a straightforward calculation. Moreover we obtain the same relation between \( \gamma \) and \( \alpha \):
		\bdm
			\frac{\Gamma_{\gamma, C}(z)_{n_0}}{2 \pi} = \frac{1}{\beta(z, \alpha)}	
		\edm
		where
		\bdm
			\beta(z, \alpha) = \lim_{\omega \rightarrow \infty}  \lambda(z-n_0 \omega, \alpha)
		\edm
		but
		\bdm
			\frac{1}{\lambda(z-n_0 \omega, \alpha)} = \frac{1}{\lambda(-i, \alpha)} - (z-n_0 \omega + i) \Big( \mathcal{G}_{-n_0 \omega +\bar{z}} (\vec{x}) , \: \mathcal{G}_{-} (\vec{x}) \Big)
		\edm
		and then
		\bdm
			\frac{1}{\beta(z, \alpha)} =  \lim_{\omega \rightarrow \infty} \bigg[ \frac{1}{\lambda(-i, \alpha)} - (z-n_0 \omega + i) \Big( \mathcal{G}_{-n_0 \omega +\bar{z}} (\vec{x}) , \: \mathcal{G}_{-} (\vec{x}) \Big) \bigg] =
		\edm
		\bdm
			= \frac{2i}{1+e^{i \alpha}} \int_0^{\infty} dk \frac{1}{|k^2+i|^2} \big| \varphi_{k0}(\vec{y}_0) \big|^2 + \int_0^{\infty} dk \frac{1}{k^2+i} \big| \varphi_{k0}(\vec{y}_0) \big|^2 +
		\edm
		\bdm
			- \int_0^{\infty} dk \frac{1}{k^2-z} \big| \varphi_{kn_0}(\vec{y}_0) \big|^2
		\edm
		\begin{flushright} 
			\( \Box \)
		\end{flushright}

\section{The Rotating Blade in 3D}

\subsection{The Hamiltonian}

Let \( D \) be the half-disc \( D \equiv \{ (r, \theta, \phi) \in \mathbb{R}^3 \: | \: 0 \leq r \leq A, \: 0 \leq \theta \leq \pi, \: \phi = 0 \} \) and \( \Theta_D(x,z) \) its characteristic function. The formal time-dependent Hamiltonian of the system is given by
	\martin
		\label{For1}
		H(t) = H_0 + \alpha(x,z) \: R(t) \: \Theta_D(x,z) \: \delta(y)
	\end{equation}
where \( R(t) \Psi(\vec{x}) = \Psi(\mathcal{R}(t)^{-1} \: \vec{x}) \) and \( \| \alpha \|_{\infty} < \infty \). Therefore in the rotating frame the formal generator of time evolution is
	\bdm
		K = H_0 - \omega J + \alpha \: \Theta_D(x,z) \: \delta(y)
	\edm
or more rigorously a self-adjoint extension of the symmetric operator
	\bdm
		K_{D} = H_{\omega}
	\edm
	\bdm
		\mathcal{D}(K_D) = C^{\infty}_0 (\mathbb{R}^3 - D)
	\edm
The Hamiltonian cannot be easily defined with the method of quadratic form, because of its unboundedness from below. Hence we shall pursue a different strategy: we shall define a sequence of cut-off Hamiltonians which converge to the operator \( H_{\omega} \) in the strong resolvent sense and that are self-adjoint and bounded from below; then we shall add the singular perturbation and prove that the so obtained operators are self-adjoint. Finally we shall prove that the limit (in the strong resolvent sense) of the sequence of cut-off perturbed Hamiltonians is a self-adjoint operator that we shall identify with the Hamiltonian of the system.
\newline
So let
	\martin
		H^L_{\omega} = H_{\omega} \: \Pi_L
	\sileno
where \( \Pi_L \) is the projector on the subspace of \( L^2(\mathbb{R}^3) \) generated by functions of the form \( \chi(r) Y_l^m(\theta, \phi) \), with \( l \leq L \). It is very easy to prove that the operator \( H^L_{\omega} \) is self-adjoint on the domain \( H^2(\mathbb{R}^3) \): the operator \( J \) is bounded on the domain of the projector \( \Pi_L \) and therefore it is an infinitesimally bounded perturbation of \( H_0 \), so that we can apply the Kato Theorem \cite{Kato1}. Moreover for each \( z \in \varrho(H_{\omega}^L) \) the resolvent \( (H_{\omega}^L - z )^{-1} \) is given by an integral operator with kernel
	\martin
		\mathcal{G}_z^L(\vec{x}, \vec{x}^{\prime}) = \int_0^{\infty} dk \sum_{l=0}^L \sum_{m=-l}^l \frac{\varphi^*_{klm}(\vec{x}^{\prime}) \: \varphi_{klm}(\vec{x})}{k^2 - m \omega - z} 
	\sileno

	\begin{pro}
		\label{Cutoff1}
		The sequence of cut-off Hamiltonians converge as \( L \rightarrow \infty \) in the strong resolvent sense to the self-adjoint operator \( H_{\omega} \).
	\end{pro}
	
	\emph{Proof:}	For each \( L \in \mathbb{N} \) and \( z \in \mathbb{C}-\mathbb{R} \), the function \( \mathcal{G}_z^L(\vec{x}, \vec{x}^{\prime}) \) belongs to \( L^2(\mathbb{R}^3, d^3\vec{x}) \):
		\bdm
			\big\| \mathcal{G}_z^L(\vec{x}, \vec{x}^{\prime}) \big\|^2 \leq \big\| \mathcal{G}_z (\vec{x}, \vec{x}^{\prime}) \big\|^2 < \infty	
		\edm
		and then the result is a straightforward consequence of Proposition \ref{Gre1}. The operator \( H_{\omega} \) was studied in \cite{Enss1, Tip1}.
		\begin{flushright} 
			\( \Box \)
		\end{flushright}

Now we can defined the perturbed cut-off Hamiltonians with the method of quadratic form: let\footnote{ Here \( d\mu_D(\vec{r}) \) stands for the restriction of the Lebesgue measure to \( D \), namely \( d\mu_D(\vec{r}) \equiv r^2 \: dr \: d\cos\theta \) for \( \vec{r} = (r, \theta) \in D \); \( \vec{r} \) denotes the restriction of \( \vec{x} \in \mathbb{R}^3 \) to \( D \), i.e. \( \vec{r} \equiv (r, \theta) \).}
	\martin
		\mathcal{F}_{\alpha, L}(\Psi, \Psi) = F_{\omega, L}(\Psi, \Psi) - \int_{D} d\mu_D(\vec{r}) \: \alpha(\vec{r}) \: \big| \Psi\big|_D(\vec{r}) \big|^2
	\sileno
where \(  F_{\omega, L} \) is the closed\footnote{The form \( F_{\omega, L} \) is closed on the domain \( \mathcal{D}(F_{\omega, L}) = H^1(\mathbb{R}^3) \).} semibounded quadratic form associated to \( H_{\omega}^L \). The form \( \mathcal{F}_{\alpha, L} \) is well defined if \( \Psi \in \mathcal{D}(F_{\omega, L}) \) and \( \alpha \) is a smooth real function on \( D \) bounded away from \( 0 \). 

	\begin{pro}
		\label{Form1}
		Let \( z \in \mathbb{C}-\mathbb{R} \), the form \( \mathcal{F}_{\alpha, L} \) can be written in the following way, 
		\martin
			\mathcal{F}_{\alpha, L}(\Psi, \Psi) = \mathcal{F}^{z}_{\omega, L}(\Psi, \Psi) + \Phi^{z}_{\alpha, L}(\xi_{\Psi}, \xi_{\Psi}) - 2 \Im(z) \: \Im \Big[ \big( \Psi \: , \:  \tilde{\mathcal{G}}^L_{z} \xi_{\Psi} \big) \Big]
		\sileno
		where
		\martin
			\mathcal{F}^{z}_{\omega, L}(\Psi, \Psi) =  F_{\omega, L}(\Psi -  \tilde{\mathcal{G}}^L_{z} \xi_{\Psi}, \Psi - \tilde{\mathcal{G}}^L_{z} \xi_{\Psi}) - \Re(z) \| \Psi -  \tilde{\mathcal{G}}^L_{z} \xi_{\Psi} \|^2 + \Re(z) \| \Psi \|^2
		\sileno
		\martin
			\Phi^z_{\alpha, L}(\xi_{\Psi}, \xi_{\Psi}) = \Re \Big[ \big( \xi_{\Psi} \: , \: \Gamma^L_{\alpha}(z) \: \xi_{\Psi} \big)_{L^2(D, d\mu_D)} \Big]
		\sileno
		and
		\martin
			\label{Gam0}
			\Big[ \Gamma^L_{\alpha}(z) \: \xi_{\Psi} \Big] (\vec{r}) = \frac{\xi_{\Psi}(\vec{r})}{\alpha(\vec{r})} - \int_D d\mu_D(\vec{r}^{\prime}) \:\: \mathcal{G}^L_{z}(\vec{x}, \vec{x}^{\prime})\big|_{\vec{x}, \vec{x}^{\prime} \in  D} \: \xi_{\Psi}(\vec{r}^{\prime})
		\sileno
		\bdm
			\big(\tilde{\mathcal{G}}^L_{z} \xi \big) (\vec{x}) \equiv \int_D d\mu_D(\vec{r}^{\prime}) \:\: \mathcal{G}^L_z(\vec{x}, \vec{x}^{\prime})\big|_{\vec{x}^{\prime} \in D} \: \xi(\vec{r}^{\prime})
		\edm
	\end{pro}

	\emph{Proof:}	The result follows from a simple calculation: setting 
		\martin
			\xi_{\Psi}(\vec{r}) = \alpha(\vec{r}) \: \Psi\big|_D(\vec{r})
		\sileno
		one has
		\bdm
			\mathcal{F}_{\alpha, L}(\Psi, \Psi) - F_{\omega, L}(\Psi -  \tilde{\mathcal{G}}^L_{z} \xi_{\Psi}, \Psi - \tilde{\mathcal{G}}^L_{z} \xi_{\Psi}) = \big( \tilde{\mathcal{G}}^L_{z} \xi \: , H^{L}_{\omega} (\Psi - \tilde{\mathcal{G}}^L_{z} \xi) \big) + 
		\edm
		\bdm
			+ \big( \Psi \: , \: H^L_{\omega} \tilde{\mathcal{G}}^L_{z} \xi \big) - \int_D d\mu_D \: \frac{|\xi_{\Psi}|^2}{\alpha} = 
		\edm
		\bdm
			= \int_D d\mu_D \: \frac{|\xi_{\Psi}|^2}{\alpha} - \big( \tilde{\mathcal{G}}^L_{z} \xi \: , \: (H_{\omega}^L - z^*) \tilde{\mathcal{G}}^L_{z} \xi \big) - z^* \big\| \tilde{\mathcal{G}}^L_{z} \xi \big\|^2 + 2 \Re \Big[ z \big( \Psi \: , \: \tilde{\mathcal{G}}^L_{z} \xi \big) \Big] = 
		\edm
		\bdm
			= \Phi^z_{\alpha, L}(\xi_{\Psi}, \xi_{\Psi}) - \Re(z) \big\| \tilde{\mathcal{G}}^L_{z} \xi \big\|^2 + 2 \Re \Big[ z \big( \Psi \: , \: \tilde{\mathcal{G}}^L_{z} \xi \big) \Big] 
		\edm
		since 
		\bdm
			\Im(z) \: \big\| \tilde{\mathcal{G}}^L_{z} \xi \big\|^2 = \Im \Big[ \big( \tilde{\mathcal{G}}^L_{z} \xi \: , \: (H_{\omega}^L - z^*) \tilde{\mathcal{G}}^L_{z} \xi \big) \Big]
		\edm
		but
		\bdm
			\big\| \tilde{\mathcal{G}}^L_{z} \xi \big\|^2 = \big\| \Psi - \tilde{\mathcal{G}}^L_{z} \xi \big\|^2 - \big\| \Psi \big\|^2 + 2 \Re \Big[ \big( \Psi \: , \: \tilde{\mathcal{G}}^L_{z} \xi \big) \Big]
		\edm
		so that we obtain the result.
		\begin{flushright} 
			\( \Box \)
		\end{flushright}

Of course the form \( \mathcal{F}_{\alpha, L} \) is independent on \( z \) and the decomposition \( \Psi = \varphi_z + \tilde{\mathcal{G}}^L_{z} \xi_{\Psi} \) is unique, since \( \tilde{\mathcal{G}}^L_{z} \xi_{\Psi} \notin \mathcal{D}(F_{\omega, L}) \) if \( \xi_{\Psi} \in L^2(D, d\mu_D) \). Moreover the form \( \Phi^z_{\alpha, L}(\xi, \xi) \) is bounded and one can choose \( z \in \mathbb{C} \) such that the form satisfies another useful inequality:
 
	\begin{pro}	
		\label{Bou1}
		The form \( \Phi^z_{\alpha, L}(\xi, \xi) \) is bounded for each \( \xi \in L^2(D, d\mu_D) \).
	\end{pro}

	\emph{Proof:}	The first term of the form is of course bounded if \( \xi \in L^2(D, d\mu_D) \) and 
		\bdm
			\bigg| \int_D d \mu_D \: \xi \: {\big( \tilde{\mathcal{G}}^L_{z} \xi \big)}^* \big|_D \bigg| \leq \big\| \xi \big\|_{L^2(D, d\mu_D)} \: \big\| \big( \tilde{\mathcal{G}}^L_{z} \xi \big) \big|_D \big\|_{L^2(D, d\mu_D)}
		\edm
		but we are going to prove that the function \( ( \tilde{\mathcal{G}}^L_{z} \xi)|_D(\vec{r}) \) is bounded \( \forall \: \vec{r} \in D \), so that
		\bdm
			\big\| \big( \tilde{\mathcal{G}}^L_{z} \xi \big) \big|_D \big\|_{L^2(D, d\mu_D)} < C(A) \: \| \xi \|^2_{L^2(D, d\mu_D)}
		\edm
		and hence the result. Indeed
		\bdm
			\Big| \big( \tilde{\mathcal{G}}^L_{z} \xi \big)\big|_D(\vec{r}) \Big|^2 = \Big| \Big( \mathcal{G}_{z}^L(\vec{x}^{\prime}, \vec{x})\big|_{\vec{x}, \vec{x}^{\prime} \in D} \: , \:  \xi(\vec{r}^{\prime}) \Big)_{L^2(D, d\mu_D)} \Big|^2 \leq
		\edm
		\bdm
			\leq \big\|  \mathcal{G}^L_{z}(\vec{x}^{\prime}, \vec{x}) \big\|^2_{L^2(D, d\mu_D(\vec{r}^{\prime}))} \: \big\| \xi \big\|^2_{L^2(D, d\mu_D)} \leq C \:\:  \big\| \xi \big\|^2_{L^2(D, d\mu_D)}
		\edm
		since the Green's function \( \mathcal{G}^L_{z}(\vec{x}, \vec{y}_0) \) belongs to \( L^2(\mathbb{R}^3) \), for each \( z \in \mathbb{C}-\mathbb{R} \) and \( \vec{y}_0 \in \mathbb{R}^3 \).
		\begin{flushright} 
			\( \Box \)
		\end{flushright}

	\begin{pro}
		\label{Ine1}
		For each smooth real function \( \alpha \) on \( D \) bounded away from \( 0 \), there exists \( \zeta \in \mathbb{R} \), \( \zeta < 0 \) such that, for each \( z \in \mathbb{C}-\mathbb{R} \), \( \Re(z) < \zeta \), the following inequality holds
		\martin
		\label{Ineq1}
			\Phi^{z}_{\alpha, L}(\xi, \xi) - 2 \Im(z) \: \Im \Big[ \big( \Psi \: , \:  \tilde{\mathcal{G}}^L_{z} \xi_{\Psi} \big) \Big] - \big( \Re(z) + \omega L \big) \: \| \Psi -  \tilde{\mathcal{G}}^L_{z} \xi_{\Psi} \|^2 > 0
		\sileno
	\end{pro}
	
	\emph{Proof:} We first point out that (see Proposition \ref{Gre1})
		\bdm
			\lim_{\Re(z) \rightarrow \infty} \big\| \mathcal{G}^L_{z}(\vec{x}, \vec{y}_0) \big\| \leq C(\Im(z)) < \infty
		\edm
		Thus, since the form \( \Phi^{z}_{\alpha, L}(\xi, \xi) \) remains bounded for each \( z \in \mathbb{C}-\mathbb{R} \), \( \Im(z) \neq 0 \), and
		\bdm
			\lim_{\Re(z) \rightarrow \infty} \Re(z) \| \Psi -  \tilde{\mathcal{G}}^L_{z} \xi_{\Psi} \|^2 = \infty
		\edm
		\bdm
			\Big| \Im(z) \: \Im \Big[ \big( \Psi \: , \:  \tilde{\mathcal{G}}^L_{z} \xi_{\Psi} \big) \Big] \Big| \leq C(\Im(z)) \: \big\| \xi \big\|^2
		\edm
		we can always found a \( \zeta \) satisfying the requirement.
		\begin{flushright} 
			\( \Box \)
		\end{flushright}

But now we can prove that the complete form \( \mathcal{F}_{\alpha, L} \) is closed and bounded from below:
	
	\begin{teo}
		\label{Forma1}
		The form \( \mathcal{F}_{\alpha, L} \) is bounded from below and closed on the domain
		\martin
			\mathcal{D}(\mathcal{F}_{\alpha, L}) = \big\{ \Psi \in L^2(\mathbb{R}^3) \: | \: \exists \xi_{\Psi} \in L^2(D, d\mu_D), \Psi - \tilde{\mathcal{G}}^L_{z} \xi_{\Psi} \in H^1(\mathbb{R}^3) \big\} 
		\sileno
		where \( z \in \mathbb{C} - \mathbb{R} \).
	\end{teo}
		
	\emph{Proof:}	Semiboundedness is trivial thanks to Proposition \ref{Ine1}: since the form \( \mathcal{F}_{\alpha, L} \) does not depend on \( z \), we can choose \( z \in \mathbb{C} - \mathbb{R} \), \( \Re(z) < \zeta \), so that the inequality (\ref{Ineq1}) applies and
		\bdm
			\mathcal{F}_{\alpha, L}(\Psi, \Psi) \geq F_{\omega, L}(\Psi -  \tilde{\mathcal{G}}^L_{z} \xi_{\Psi}, \Psi - \tilde{\mathcal{G}}^L_{z} \xi_{\Psi}) + \omega L \: \| \Psi -  \tilde{\mathcal{G}}^L_{z} \xi_{\Psi} \|^2 + \Re(z) \: \| \Psi \|^2 \geq 
		\edm
		\bdm
			\geq  F_{0}(\Psi -  \tilde{\mathcal{G}}^L_{z} \xi_{\Psi}, \Psi - \tilde{\mathcal{G}}^L_{z} \xi_{\Psi}) + \Re(z) \: \| \Psi \|^2 \geq \Re(z) \: \| \Psi \|^2
		\edm
		So it remains to prove closure. Let \( \Psi_n = \varphi_n + \tilde{\mathcal{G}}_z \xi_n \) be a sequence in \( \mathcal{D}(\mathcal{F}_{\alpha, L}) \) converging to \( \Psi \) in the norm topology of \( L^2(\mathbb{R}^3) \), such that\footnote{\( F_0 \) is simply the form associated to the free Hamiltonian, i.e. \( F_0 (\Psi, \Psi) = \int | \nabla \Psi |^2 \).}
		\bdm
			\lim_{n,m \rightarrow \infty} \big( \mathcal{F}_{\alpha, L} - \Re(z) \big) (\Psi_n - \Psi_m) = 0
		\edm
		\bdm
			\lim_{n,m \rightarrow \infty} \big( \mathcal{F}_{\alpha, L} - \Re(z) \big) (\Psi_n - \Psi_m) \geq \lim_{n,m \rightarrow \infty} F_{0} (\varphi_n - \varphi_m) \geq 0
		\edm
		so that
		\bdm
			\lim_{n,m \rightarrow \infty} F_0 (\varphi_n - \varphi_m) = 0 
		\edm
		and
		\bdm	
			\lim_{n,m \rightarrow \infty} \Phi_{\alpha, L}^z (\xi_n - \xi_m) = 0
		\edm
		The result easily follows, because \( F_0 \) and \( \Phi_{\alpha, L}^z \) are closed forms (see Proposition \ref{Bou1}).
		\begin{flushright} 
			\( \Box \)
		\end{flushright}

Thus the form \( \mathcal{F}_{\alpha, L} \) defines a semibounded self-adjoint operator:
	
	\begin{pro}
		\label{CutResolvent1}
		The operators \( K^L_{\alpha} \) defined below are self-adjoint:
		\bdm
			\mathcal{D}(K^L_{\alpha}) = \big\{ \Psi \in L^2(\mathbb{R}^3) \: | \: \exists \xi_{\Psi} \in L^2(D, d\mu_D), \Psi - \tilde{\mathcal{G}}^L_z \xi_{\Psi} \in \mathcal{D}(H^L_{\omega}),
		\edm
		\martin
			\big( \Psi - \tilde{\mathcal{G}}^L_z \xi_{\Psi} \big)\big|_D = \Gamma^L_{\alpha}(z) \xi_{\Psi} \big\}
		\sileno
		\martin
		\label{Cut1}
			\big( K^L_{\alpha} - z \big) \Psi = \big( H^L_{\omega} - z \big) \big( \Psi - \tilde{\mathcal{G}}^L_z \xi_{\Psi} \big)
		\sileno
		where \( \alpha \in \mathrm{C}(D) \), \( \alpha(\vec{r}) \neq 0 \), for each \( \vec{r} \in D \).
		\newline
		Moreover
		\bdm
			\Big[ \big( K^L_{\alpha} - z \big)^{-1} \Psi \Big](\vec{x}) = \Big[ \big( H^L_{\omega} - z \big)^{-1} \Psi \Big] (\vec{x}) \: + 
		\edm
		\martin
		\label{CutRes1}
			+ \int_D d^2\vec{r}^{\prime} \:\:  \big[ \Gamma^L_{\alpha}(z) \big]^{-1} \Big[ \big( H^L_{\omega} - z \big)^{-1} \Psi \Big]\Big|_D (\vec{r}^{\prime}) \: \mathcal{G}^L_z(\vec{x}, \vec{x}^{\prime})\big|_{\vec{x}^{\prime} \in D}
		\sileno\
		for each \( z \in\varrho(K_{\alpha}) \).
	\end{pro}
	
	\emph{Proof:}	The result easily follows from Theorem \ref{Forma1}. The explicit expression of the resolvent is a direct consequence of  the equation (\ref{Cut1}). We want only to remark that the operator \( \Gamma^L_{\alpha}(z) \) is invertible if \( \Im(z) \neq 0 \): the form \( \Phi_{\alpha, L}^z \) can be written in the following way
		\bdm
			\Phi^z_{\alpha, L}(\xi, \xi) \equiv \int_D d\mu_D \:\: \frac{|\xi|^2}{\alpha} - \Re(z) \big\| \tilde{\mathcal{G}}^L_z \xi \big\|^2
		\edm
		Since \( \big\| \tilde{\mathcal{G}}^L_z \xi \big\|^2 \) is bounded by \( C(\Im(z)) \: \| \xi \|^2 \), if \( \Im(z) \neq 0 \), we can always choose the real part of \( z \) is such a way that the form is positive.
		\begin{flushright} 
			\( \Box \)
		\end{flushright}

At last we can remove the cut-off in the angular momentum and define the Hamiltonian of the system:
	
	\begin{teo}
		\label{Resolvent1}
		For each \( \alpha \in \mathrm{C}(D) \), \( \alpha(\vec{r}) \neq 0 \), \( \forall \: \vec{r} \in D \), the sequence of semibounded self-adjoint operators \( K_{\alpha}^L \) converge as \( L \rightarrow \infty \) in the strong resolvent sense to the self-adjoint (unbounded from below) operator \( K_{\alpha} \):
		\bdm
			\mathcal{D}(K_{\alpha}) = \big\{ \Psi \in L^2(\mathbb{R}^3) \: | \: \exists \xi_{\Psi} \in L^2(D, d\mu_D), \Psi - \tilde{\mathcal{G}}_z \xi_{\Psi} \in \mathcal{D}(H_{\omega}),
		\edm
		\martin
			\big( \Psi - \tilde{\mathcal{G}}_z \xi_{\Psi} \big)\big|_D = \Gamma_{\alpha}(z) \xi_{\Psi} \big\}
		\sileno
		\martin
		\label{Ham3}
			\big( K_{\alpha} -z \big) \Psi = \big( H_{\omega} - z \big) \big( \Psi - \tilde{\mathcal{G}}_z \xi_{\Psi} \big)
		\sileno
		where
		\martin
			\label{Gam1}
			\Big[ \Gamma_{\alpha}(z) \: \xi_{\Psi} \Big] (\vec{r}) = \frac{\xi_{\Psi}(\vec{r})}{\alpha(\vec{r})} - \int_D d\mu_D(\vec{r}^{\prime}) \:\: \mathcal{G}_{z}(\vec{x}, \vec{x}^{\prime})\big|_{\vec{x}, \vec{x}^{\prime} \in  D} \: \xi_{\Psi}(\vec{r}^{\prime})
		\sileno
		\bdm
			\big(\tilde{\mathcal{G}}_{z} \xi \big) (\vec{x}) \equiv \int_D d\mu_D(\vec{r}^{\prime}) \:\: \mathcal{G}_z(\vec{x}, \vec{x}^{\prime})\big|_{\vec{x}^{\prime} \in D} \: \xi(\vec{r}^{\prime})
		\edm
		Moreover the resolvent of \( K_{\alpha} \) is			
		\bdm
			\Big[ \big( K_{\alpha} - z \big)^{-1} \Psi \Big](\vec{x}) = \Big[ \big( H_{\omega} - z \big)^{-1} \Psi \Big] (\vec{x}) \: + 
		\edm
		\martin
			\label{Resolvent3}
			+ \int_D d^2\vec{r}^{\prime} \:\:  \Gamma^{-1}_{\alpha}(z) \Big[ \big( H_{\omega} - z \big)^{-1} \Psi \Big]\Big|_D (\vec{r}^{\prime}) \: \mathcal{G}_z(\vec{x}, \vec{x}^{\prime})\big|_{\vec{x}^{\prime} \in D}
		\sileno
		for each \( z \in\varrho(K_{\alpha}) \).
	\end{teo}

	\emph{Proof:}	The key point of the proof is the application of the Trotter-Kato Theorem (see Theorem VIII.22 in \cite{Reed2}) to the sequence of self-adjoint operators \( K_{\alpha}^L \): we shall prove that \( (K_{\alpha}^L - z)^{-1} \) converge in the strong sense for all \( z \in \mathbb{C}- \mathbb{R} \) to the operator \( (K_{\alpha} - z)^{-1} \), then the Trotter-Kato Theorem guarantees that there exists a self-adjoint operator \( T \) such that \( K_{\alpha}^L \) converges in the strong resolvent sense to \( T \). The identification of \( T \) with \( K_{\alpha} \) is then trivial.
		\newline
		So we shall start with the analysis of the sequence of bounded operators \linebreak \( (K_{\alpha} - z)^{-1} \), \( z \in \mathbb{C} - \mathbb{R} \), defined in (\ref{CutRes1}): thanks to Proposition \ref{Cutoff1}, the first part of the resolvent converges in the strong sense to \( (H_{\omega} - z)^{-1} \), so that, in order to prove convergence of the whole operator, we need to consider the second part,
		\bdm
			\int_D d^2\vec{r}^{\prime} \:\:  \big[ \Gamma^L_{\alpha}(z) \big]^{-1} \Big[ \big( H^L_{\omega} - z \big)^{-1} \Psi \Big]\Big|_D (\vec{r}^{\prime}) \: \mathcal{G}^L_z(\vec{x}, \vec{x}^{\prime})\big|_{\vec{x}^{\prime} \in D}
		\edm
		but, for the same reason, 
		\bdm
			\lim_{L \rightarrow \infty} \mathcal{G}^L_z(\vec{x}, \vec{x}^{\prime})\big|_{\vec{x}^{\prime} \in D} =  \mathcal{G}_z(\vec{x}, \vec{x}^{\prime})\big|_{\vec{x}^{\prime} \in D}
		\edm
		in \( L^2(\mathbb{R}^3) \) and
		\bdm
			\lim_{L \rightarrow \infty} \Big[ \big( H^L_{\omega} - z \big)^{-1} \Psi \Big]\Big|_D (\vec{r}^{\prime}) =  \Big[ \big( H_{\omega} - z \big)^{-1} \Psi \Big]\Big|_D (\vec{r}^{\prime})
		\edm
		in \( L^2(D, d\mu_D) \), for all \( \Psi \in L^2(\mathbb{R}^3) \). Hence, to complete the first part of the proof, it is sufficient to show that
		\bdm
			\lim_{L \rightarrow \infty}  \big[ \Gamma^L_{\alpha}(z) \big]^{-1} =  \Gamma_{\alpha}^{-1}(z) 
		\edm
		in the norm topology of \( L^2(D, d\mu_D) \), but this is again a consequence of Proposition \ref{Cutoff1}: for each \( L \) the operator \( \Gamma^L_{\alpha}(z) \) is invertible (see the Proof of Proposition \ref{CutResolvent1}) and, in the same way, we can prove that \(  \Gamma_{\alpha}^{-1}(z) \) is bounded and well defined, if \( \Im(z) \neq 0 \); moreover it is easy to see that 
		\bdm
			\lim_{L \rightarrow \infty}  \Gamma^L_{\alpha}(z) = \Gamma_{\alpha}(z) 
		\edm
		We have then proved that, for each \( z \in \mathbb{C} - \mathbb{R} \),
		\bdm
			\mathrm{s-}\lim_{L \rightarrow \infty} \big( K^L_{\alpha} - z \big)^{-1} = \big( K_{\alpha} - z \big)^{-1}
		\edm
		and the operator \( (K_{\alpha} - z)^{-1} \) has of course a dense range. Thus the Trotter-Kato Theorem applies and the limiting self-adjoint operator \( T \) is immediately identified with \( K_{\alpha} \): the domain of \( K_{\alpha} \) is given by functions of the form \( (K_{\alpha} - z)^{-1} \Psi \), \( \Psi \in L^2(\mathbb{R}^3) \), and the action of the operator on its domain follows from (\ref{Resolvent3}).
		\begin{flushright} 
			\( \Box \)
		\end{flushright}	

	\begin{teo}
		\label{Spe3}
		The spectrum of \( K_{\alpha} \) is purely absolutely continuous and
		\bdm
			\sigma(K_{\alpha}) = \sigma_{\mathrm{ac}} (K_{\alpha}) = \sigma(H_{\omega}) = \mathbb{R}
		\edm
	\end{teo}

	\emph{Proof:} First of all we shall prove that the operator 
		\bdm
			\mathcal{R}_{\alpha}^z \equiv \big( K_{\alpha} - z \big)^{-1} - \big( H_{\omega} - z \big)^{-1} 
		\edm
		is a compact operator \( \forall \: z \in \mathbb{C} - \mathbb{R} \). Let \( \Psi_n \) be a weakly convergent sequence in \( L^2(\mathbb{R}^3) \), namely \( (\varphi \: , \Psi_n - \Psi_m) \rightarrow 0 \) when \( n, m \rightarrow \infty \) for each \( \varphi \in L^2(\mathbb{R}^3) \),
		\bdm
			\mathcal{R}_{\alpha}^z \: (\Psi_n - \Psi_m) = \int_D d^2\vec{r}^{\prime} \:\:  \Gamma^{-1}_{\alpha}(z) \Big[ \big( H_{\omega} - z \big)^{-1} (\Psi_n - \Psi_m) \Big]\Big|_D \: \mathcal{G}_z(\vec{x}, \vec{x}^{\prime})\big|_{\vec{x}^{\prime} \in D}
		\edm
		and
		\bdm
			\big\| \mathcal{R}_{\alpha}^z \: (\Psi_n - \Psi_m) \big\| \leq \big\| \mathcal{G}_z \big\| \big\| \Gamma_{\alpha}^{-1}(z) \big\| \: \Big| \Big( \mathcal{G}^*_{z^*} \: , \: \Psi_n - \Psi_m \Big) \Big| \leq
		\edm
		\bdm
			\leq C  \: \Big| \Big( \mathcal{G}_{z^*} \: , \: \Psi_n - \Psi_m \Big) \Big| \underset{n,m \rightarrow \infty}{\longrightarrow} 0
		\edm
		since the operator \( \Gamma_{\alpha}^{-1}(z) \) is bounded (see the Proof of Theorem \ref{Resolvent1}).
		\newline	
		Therefore we can apply Weyl's theorem and thus 
		\bdm
			\sigma_{\mathrm{ess}}(K_{\alpha}) = \sigma_{\mathrm{ess}}(H_{\omega}) = \mathbb{R}
		\edm
		To prove that the singular and pure points spectrum of \( K_{\alpha} \) are empty, we refer again to the limiting absorption principle. To show that the condition of the principle is satisfied, we have to consider the scalar product (where \( z = x + i \varepsilon \))
		\bdm
			\Big| \Big( \Psi \: , \mathcal{R}_{\alpha}^z \Psi \Big) \Big| = \bigg| \int_D d^2\vec{r}^{\prime} \:\:  \Gamma^{-1}_{\alpha}(z) \Big[ \big( H_{\omega} - z \big)^{-1} \Psi \Big]\Big|_D \: \Big( \Psi \: , \: \mathcal{G}_z(\vec{x}, \vec{x}^{\prime})\big|_{\vec{x}^{\prime} \in D} \Big) \bigg| \leq
		\edm
		\bdm
			\leq \big\| \Gamma_{\alpha}^{-1}(z) \big\| \bigg| \: \int_D d^2\vec{r}^{\prime} \:\: \Big( \mathcal{G}_{z^*}(\vec{x}, \vec{x}^{\prime})\big|_{\vec{x}^{\prime} \in D} \: , \: \Psi \Big) \: \Big( \Psi \: , \: \mathcal{G}_z(\vec{x}, \vec{x}^{\prime})\big|_{\vec{x}^{\prime} \in D} \Big) \bigg| 
		\edm
		The operator \( \Gamma_{\alpha}^{-1}(z) \) remains bounded when \( \varepsilon \rightarrow 0 \) and, applying the same trick used in the Proof of Theorem \ref{Spe1}, one has
		\bdm	
			\lim_{\varepsilon \rightarrow 0} \Big( \mathcal{G}_{x-i\varepsilon}(\vec{x}, \vec{x}^{\prime})\big|_{\vec{x}^{\prime} \in D} \: , \: \Psi \Big) \: \Big( \Psi \: , \: \mathcal{G}_{x + i\varepsilon}(\vec{x}, \vec{x}^{\prime})\big|_{\vec{x}^{\prime} \in D} \Big) = \big| \varphi(\vec{r}^{\prime}) \big|^2 < \infty
		\edm
		where \( \Psi = (H_{\omega} - x) \varphi \) and \( \varphi \in \mathcal{D}(H_{\omega}) \), so that
		\bdm
			\sup_{0 < \varepsilon < 1} \int_a^b dx \:\: \Big| \Big( \Psi \: , \: \mathcal{R}^{x+i\varepsilon}_{\alpha} \Psi \Big) \Big|^p < \infty
		\edm
		for some \( p > 1 \) and for each interval \( [a,b] \subset \mathbb{R} \).
		\begin{flushright} 
			\( \Box \)
		\end{flushright}
		
\subsection{Asymptotic Limit of Rapid Rotation}

In this Section we shall study the asymptotic limit of rapid rotation of the unitary group
	\bdm
		U_{\mathrm{inert}}(t,s) = R(t) \: U_{\mathrm{rot}}(t-s) \: R^{\dagger}(s)
	\edm
which represents the time evolution in the inertial frame associated to the formal time-dependent Hamiltonian defined in (\ref{For1}), while \(  U_{\mathrm{rot}}(t-s) \) is the unitary group associated to the self-adjoint generator \( K_{\alpha} \): our main goal will be the proof of the following result,
	\bdm
		\mathrm{s-}\lim_{\omega \rightarrow \infty} U_{\mathrm{inert}}(t,s) = e^{-iH_{\alpha} (t-s)}
	\edm
where \( H_{\alpha} \) is the self-adjoint generator\footnote{The operator \( H_{\alpha} \) is easily defined with the method of quadratic form (see for example \cite{Reed2}): since the potential \( \alpha(r) \) is bounded, it is associated to a form infinitesimally bounded w.r.t. the free Hamiltonian \( H_0 \). Hence the operator \( H_0 + \alpha(r) \: \Theta_{D}(\vec{r}) \) is self-adjoint on the domain of \( H_0 \).} 
	\martin
		H_{\alpha} = H_0 - \alpha(\vec{r}) \: \Theta_{S}(\vec{r})
	\sileno	 
and \(  \Theta_S(\vec{r}) \) is the characteristic function of a sphere \( S \) of radius \( A \) centered at the origin. 
	
	\begin{teo}
		For every \( t,s \in \mathbb{R} \),
		\bdm
			\mathrm{s-}\lim_{\omega \rightarrow \infty} U_{\mathrm{inert}}(t,s) = e^{-iH_{\alpha} (t-s)}
		\edm
		where 
		\bdm
			H_{\alpha} = H_0 - \alpha(\vec{r}) \: \Theta_{S}(\vec{r})
		\edm
	\end{teo}

	\emph{Proof:}	See the Proof of Theorem \ref{Asy1} and the following Lemma \ref{Con3}.
		\begin{flushright} 
			\( \Box \)
		\end{flushright}

	\begin{lem}
		\label{Con3}
		For every \( z \in \mathbb{C} \), \( \Im(z) > 0 \),
		\bdm
			\mathrm{s-}\lim_{\omega \rightarrow \infty} \int_{-\infty}^0 dt \: e^{-izt} \: U^{*}_{\mathrm{inert}}(t,0) = -i \big(H_{\alpha} - z \big)^{-1}
		\edm
	\end{lem}
		
	\emph{Proof:}	Like in the Proof of Lemma \ref{Con1}, we shall prove the result on the dense subset of \( L^2(\mathbb{R}^3) \) given by functions of the form \( \Psi(\vec{x}) = \chi(r) Y_{l_0}^{m_0}(\theta, \phi) \), with \( l_0 = 0, \ldots \infty \) and \( m_0 = -l_0, \ldots, l_0 \). The first part of the Proof of Lemma \ref{Con1} still applies, so that it is sufficient to prove that  
		\bdm
			\lim_{\omega \rightarrow \infty} \big( K_{\alpha}+m_0 \omega - z \big)^{-1} \Psi(\vec{x}) =  \big(H_{\alpha} - z \big)^{-1} \Psi(\vec{x})
		\edm
		First of all we observe that
		\bdm
			 \big( K_{\alpha} + m_0 \omega - z \big)^{-1} \Psi = \big( H_{\omega} + m_0 \omega - z \big)^{-1} \Psi \: +
		\edm
		\bdm
			+ \: \Big(  \Gamma^{-1}_{\alpha}(z^* - m_0 \omega) \Big[ \big( H_{\omega} + m_0 \omega - z^* \big)^{-1} \Psi \Big]\Big|_D \: , \:  \mathcal{G}_{z - m_0 \omega}(\vec{x}, \vec{x}^{\prime})\big|_{\vec{x}^{\prime} \in D} \Big)_{L^2(D, d\mu_D)}
		\edm
		and
		\bdm
			\lim_{\omega \rightarrow \infty} \big( H_{\omega} + m_0 \omega - z \big)^{-1} \Psi = \big(H_0 - z \big)^{-1} \Psi
		\edm
		as we have proved in Lemma \ref{Con1}.
		\newline
		Therefore we need only to study the second part of the resolvent: it is easy to see that 
		\bdm
			\lim_{\omega \rightarrow \infty} \Big[ \big( H_{\omega} + m_0 \omega - z \big)^{-1} \Psi \Big]\Big|_D = \Big[ \big(H_0 - z \big)^{-1} \Psi \Big]\Big|_D
		\edm
		in \( L^2(D, d\mu_D) \). Moreover, since \( \big[ \big( H_0 - z \big)^{-1} \Psi \big]\big|_D (\vec{r}) \) is a function of the form \( \chi(r) Y_{l_0}^{m_0}(\theta, 0) \), we can apply the result found in the following Lemma \ref{Gamma1}:
		\bdm
			\lim_{\omega \rightarrow \infty} \Gamma^{-1}_{\alpha}(z - m_0 \omega) \Big[ \big( H_{\omega} + m_0 \omega - z \big)^{-1} \Psi \Big]\Big|_D = 
		\edm
		\bdm
			= \Xi_{\alpha}(z) \Big[ \big( H_0 - z \big)^{-1} \Psi \Big]\Big|_D =
		\edm
		\bdm
			= \alpha(\vec{r}) \Theta_D(\vec{r}) \big( H_0 - \alpha(\vec{r}) \: \Theta_S(\vec{r}) - z \big)^{-1} \Psi
		\edm
		In conclusion we obtain
		\bdm
			\lim_{\omega \rightarrow \infty} \big( K_{\alpha}+m_0 \omega - z \big)^{-1} \Psi = \big( H_0 - z \big)^{-1} \Big[ 1 + \alpha \Theta_D \big( H_0 - \alpha \: \Theta_S - z \big)^{-1} \Big] \Psi =
		\edm
		\bdm
			= \big( H_0 - \alpha \: \Theta_S - z \big)^{-1} \Psi
		\edm
		\begin{flushright} 
			\( \Box \)
		\end{flushright}

	\begin{lem}
		\label{Gamma1}
		Let \( \Gamma_{\alpha}(z) \) the operator defined in (\ref{Gam1}) and \( \Psi(\vec{x}) \in L^2(\mathbb{R}^3) \) of the form \( \Psi(\vec{x}) = \chi(r) Y_{l_0}^{m_0}(\theta, \phi) \),
		\bdm
			\lim_{\omega \rightarrow \infty} \Gamma^{-1}_{\alpha}(z - m_0 \omega) \: \Psi|_D = \Xi_{\alpha}(z) \: \Psi|_D
		\edm
		in \( L^2(D, d\mu_D) \), where
		\martin
			\big( \Xi_{\alpha}(z) \Psi|_D \big) (\vec{r}) \equiv \Big[ \alpha(\vec{r}) \big( H_0 - \alpha(\vec{r}) \: \Theta_S(\vec{r}) - z \big)^{-1} \big( H_0 -z \big) \: \Psi|_D \Big] (\vec{r})
		\sileno
	\end{lem}

	\emph{Proof:}	First of all we are going to prove that
		\bdm
			\mathrm{norm-}\lim_{\omega \rightarrow \infty} \Gamma_{\alpha}(z - m_0 \omega) = \Lambda_{\alpha}(z)
		\edm
		where
		\bdm
			\big( \Lambda_{\alpha}(z) \: \xi \big) = \frac{\xi}{\alpha} - \int_D d\mu_D(\vec{r}^{\prime}) \:\: G_z^{m_0}(\vec{x},\vec{x}^{\prime})\big|_{\vec{x}, \vec{x}^{\prime} \in D} \xi(\vec{r}^{\prime})
		\edm
		for the definition of \( G_z^{m_0} \) see Proposition \ref{ReH1}.
		\newline
		Indeed
		\bdm
			\Gamma_{\alpha}(z - m_0\omega) = \Lambda_{\alpha}(z) + R_z^{m_0}
		\edm
		where \( R^{m_0}_z \) is a bounded integral operator on \( L^2(D, d\mu_D) \) with kernel
		\bdm
			R^{m_0}_z(\vec{r}, \vec{r}^{\prime}) \equiv \int_0^{\infty} \sum_{l=0}^{\infty} \underset{m \neq m_0}{\sum_{m=-l}^l} \frac{\varphi_{klm}(\vec{r}) \: \varphi_{klm}(\vec{r}^{\prime})}{k^2 - (m - m_0) \omega - z}
		\edm
		that goes to \( 0 \) when \( \omega \rightarrow \infty \) (see the Proof of Lemma \ref{Con1}).
		\newline
		Moreover \( \forall \: \omega \in \mathbb{R}^+ \) the operator \( \Gamma_{\alpha}(z) \) is invertible if \( \Im(z) \neq 0 \) (see the Proof of Theorem \ref{Resolvent1}) and, for each \( l_0 \in \mathbb{N} \), \( m_0 = -l_0, \ldots, l_0  \), \( z \in \mathbb{C} - \mathbb{R} \) it can be seen that the operator \( \Lambda_{\alpha} \) is also invertible: indeed, let \( \Psi \) is the dense subset of \( L^2(D, d\mu_D) \) given by functions of the form \( \chi(r) Y_{l_0}^{m_0}(\theta, 0) \), 
		\bdm
			\big( \Lambda_{\alpha}(z) \: \Psi|_D \big) (r, \theta) = \frac{\Psi|_D}{\alpha} (r, \theta) - \frac{Y_{l_0}^{m_0}(\theta, 0)}{2 \pi} \int_0^A dr^{\prime} \: {r^{\prime}}^2 \:\: g_z^{l_0}(\vec{r},\vec{r}^{\prime}) \: \chi(\vec{r}^{\prime})
		\edm
		and 
		\bdm
			\Big[ \big(H_0 - z \big) \Lambda_{\alpha}(z) \: \Psi|_D \Big] (\vec{r}) = \bigg[ \big(H_0 - z \big) \frac{\Psi|_D}{\alpha} \bigg] (\vec{r}) - \Theta_D(\vec{r}) \: \Psi|_D(\vec{r})
		\edm
		so that \( \Lambda^{-1}_{\alpha}(z) \Psi|_D = \Xi_{\alpha}(z) \Psi|_D \). 
		\begin{flushright} 
			\( \Box \)
		\end{flushright}

\section{The Rotating Blade in 2D}

\subsection{The Hamiltonian}

The formal time-dependent Hamiltonian of the system is given by the operator
	\martin
		H(t) = H_0 + \alpha(x) \: R(t) \: \Theta_A(x) \: \delta(y)
	\end{equation}
where \( \Theta_A(x) \) is the characteristic function of the segment \( 0 \leq x \leq A \). In the rotating frame the generator of time evolution is a self-adjoint extension of the symmetric operator
	\bdm
		K_{S} = H_{\omega}
	\edm
	\bdm
		\mathcal{D}(K_S) = C^{\infty}_0 (\mathbb{R}^2 - S)
	\edm
where \( S \) is the segment \( S \equiv \{ (x, 0) \in \mathbb{R}^2 \: | \: 0 \leq x \leq A \} \).
\newline
In order to rigorously define the self-adjoint extensions of the operator \( K_S \), we shall proceed like in the 3D case, namely we shall introduce a sequence of cut-off perturbed Hamiltonians and then we shall identify their limit with the Hamiltonian of the system.
\newline
So let
	\martin
		H^N_{\omega} = H_{\omega} \: \Pi_N
	\sileno
where \( \Pi_N \) is the projector on the subspace of \( L^2(\mathbb{R}^2) \) generated by functions of the form \( \chi(r) e_n(\theta) \), with \( |n| \leq N \). The operator \( H^N_{\omega} \) is self-adjoint on the domain \( H^2(\mathbb{R}^2) \) (see the discussion at the beginning of Section 4) and, for each \( z \in \varrho(H_{\omega}^N) \), the resolvent \( (H_{\omega}^N - z )^{-1} \) is given by an integral operator with kernel
	\martin
		\mathcal{G}_z^N(\vec{x}, \vec{x}^{\prime}) = \int_0^{\infty} dk \sum_{n=-N}^N \frac{\varphi^*_{kn}(\vec{x}^{\prime}) \: \varphi_{kn}(\vec{x})}{k^2 - \omega n - z} 
	\sileno
	
	\begin{pro}
		\label{Cutoff2}
		The sequence of cut-off Hamiltonians converge as \( N \rightarrow \infty \) in the strong resolvent sense to the self-adjoint operator \( H_{\omega} \).
	\end{pro}
	
	\emph{Proof:}	See the Proof of Proposition \ref{Cutoff1} and Proposition \ref{Gre2}.
		\begin{flushright} 
			\( \Box \)
		\end{flushright}

The perturbed cut-off Hamiltonian is associated to the form
	\martin
		\mathcal{F}_{\alpha, N}(\Psi, \Psi) = F_{\omega, N}(\Psi, \Psi) - \int_S d\mu_S \: \alpha(r) \: \big| \Psi\big|_S(r) \big|^2
	\sileno
which is well defined\footnote{In the 2D case, the measure \( d\mu_S \) is given by \( r \: dr \).} if \( \Psi \in \mathcal{D}(F_{\omega, N}) \), \( F_{\omega, N} \) being the closed semibounded form associated to the self-adjoint operator \( H^N_{\omega} \), and \( \alpha \in C(S) \), \( \alpha(r) \neq 0 \), \( \forall r \in S \). 

	\begin{pro}
		Let \( z \in \mathbb{C}-\mathbb{R} \), the form \( \mathcal{F}_{\alpha, N} \) can be written in the following way, 
		\martin
			\mathcal{F}_{\alpha, N}(\Psi, \Psi) = \mathcal{F}^{z}_{\omega, N}(\Psi, \Psi) + \Phi^{z}_{\alpha, N}(\xi_{\Psi}, \xi_{\Psi}) - 2 \Im(z) \: \Im \Big[ \big( \Psi \: , \:  \tilde{\mathcal{G}}^N_{z} \xi_{\Psi} \big) \Big]
		\sileno
		where
		\martin
			\mathcal{F}^{z}_{\omega, N}(\Psi, \Psi) = F_{\omega, N}(\Psi -  \tilde{\mathcal{G}}^N_{z} \xi_{\Psi}, \Psi - \tilde{\mathcal{G}}^N_{z} \xi_{\Psi}) - \Re(z) \| \Psi -  \tilde{\mathcal{G}}^N_{z} \xi_{\Psi} \|^2 + \Re(z) \| \Psi \|^2
		\sileno
		\martin
			\Phi^z_{\alpha, N}(\xi_{\Psi}, \xi_{\Psi}) = \Re \Big[ \big( \xi_{\Psi} \: , \: \Gamma^N_{\alpha}(z) \: \xi_{\Psi} \big)_{L^2(S, d\mu_S)} \Big]
		\sileno
		and
		\martin
		\label{Gam2}
			\Big[ \Gamma_{\alpha}^N(z) \: \xi_{\Psi} \Big] (r) = \frac{\xi_{\Psi}(r)}{\alpha(r)} - \int_S d\mu_S(r^{\prime}) \:\: \mathcal{G}^N_{z}(\vec{x}, \vec{x}^{\prime})\big|_{\vec{x}, \vec{x}^{\prime} \in S} \: \xi_{\Psi}(r^{\prime})
		\sileno
		\bdm
			\big(\tilde{\mathcal{G}}^N_{z} \xi \big) (\vec{x}) \equiv \int_S d\mu_S(r^{\prime}) \:\: \mathcal{G}_z^N(\vec{x}, \vec{x}^{\prime})\big|_{\vec{x}^{\prime} \in S} \: \xi(r^{\prime})
		\edm
	\end{pro}

	\emph{Proof:}	See the Proof of Proposition \ref{Form1}.
		\begin{flushright} 
			\( \Box \)
		\end{flushright}

Now we shall prove that the properties of the form \( \Phi^z_{\alpha, N} \) still hold:
	
	\begin{pro}	
		\label{Bou2}
		The form \( \Phi^z_{\alpha, N}(\xi, \xi) \) is bounded for each \( \xi \in L^2(S, d\mu_S) \).
	\end{pro}

	\emph{Proof:}	Using the result proved in Proposition \ref{Gre2}, we can follow the Proof of Proposition \ref{Bou1}.
		\begin{flushright} 
			\( \Box \)
		\end{flushright}

	\begin{pro}
		\label{Ine2}
		For each smooth real function \( \alpha \) on \( S \) bounded away from \( 0 \), there exists \( \zeta \in \mathbb{R} \), \( \zeta < 0 \) such that, for each \( z \in \mathbb{C}-\mathbb{R} \), \( \Re(z) < \zeta \), the following inequality holds
		\bdm
			\Phi^{z}_{\alpha, N}(\xi, \xi) - 2 \Im(z) \: \Im \Big[ \big( \Psi \: , \:  \tilde{\mathcal{G}}^N_{z} \xi_{\Psi} \big) \Big] - \big( \Re(z) + \omega N \big) \: \| \Psi -  \tilde{\mathcal{G}}^N_{z} \xi_{\Psi} \|^2 > 0
		\edm
	\end{pro}
	
	\emph{Proof:}	See the Proof of Proposition \ref{Ine1} and Proposition \ref{Gre2}.
		\begin{flushright} 
			\( \Box \)
		\end{flushright}

We can now state the following Theorem,

	\begin{teo}
		\label{Forma2}
		The form \( \mathcal{F}_{\alpha, N} \) is bounded from below and closed on the domain
		\martin
			\mathcal{D}(\mathcal{F}_{\alpha, N}) = \big\{ \Psi \in L^2(\mathbb{R}^2) \: | \: \exists \xi_{\Psi} \in L^2(S, r dr), \Psi - \tilde{\mathcal{G}}^N_z \xi_{\Psi} \in H^1(\mathbb{R}^2) \big\} 
		\sileno
	\end{teo}
		
	\emph{Proof:} See the Proof of Theorem \ref{Forma1}.
		\begin{flushright} 
			\( \Box \)
		\end{flushright}

	\begin{pro}
		\label{CutResolvent2}
		The operators \( K^N_{\alpha} \) defined below are self-adjoint:
		\bdm
			\mathcal{D}(K^N_{\alpha}) = \big\{ \Psi \in L^2(\mathbb{R}^2) \: | \: \exists \xi_{\Psi} \in L^2(S, d\mu_S), \Psi - \tilde{\mathcal{G}}^N_z \xi_{\Psi} \in \mathcal{D}(H^N_{\omega}),
		\edm
		\martin
			\big( \Psi - \tilde{\mathcal{G}}^N_z \xi_{\Psi} \big)\big|_D = \Gamma^N_{\alpha}(z) \xi_{\Psi} \big\}
		\sileno
		\martin
		\label{Cut2}
			\big( K^N_{\alpha} - z \big) \Psi = \big( H^N_{\omega} - z \big) \big( \Psi - \tilde{\mathcal{G}}^N_z \xi_{\Psi} \big)
		\sileno
		where \( \alpha \in \mathrm{C}(D) \), \( \alpha(\vec{r}) \neq 0 \), for each \( \vec{r} \in D \).
		\newline
		Moreover
		\bdm
			\Big[ \big( K^N_{\alpha} - z \big)^{-1} \Psi \Big](\vec{x}) = \Big[ \big( H^N_{\omega} - z \big)^{-1} \Psi \Big] (\vec{x}) \: + 
		\edm
		\martin
		\label{CutRes2}
			+ \int_D d^2\vec{r}^{\prime} \:\:  \big[ \Gamma^N_{\alpha}(z) \big]^{-1} \Big[ \big( H^N_{\omega} - z \big)^{-1} \Psi \Big]\Big|_D (\vec{r}^{\prime}) \: \mathcal{G}^N_z(\vec{x}, \vec{x}^{\prime})\big|_{\vec{x}^{\prime} \in D}
		\sileno\
		for each \( z \in\varrho(K_{\alpha}) \).
	\end{pro}
	
	\emph{Proof:}	The result follows from Theorem \ref{Forma2}. Like in the 3D case it is possible to prove that the operator \( \Gamma_{\alpha}^N(z) \) is invertible if \( \Im(z) \neq 0 \).
		\begin{flushright} 
			\( \Box \)
		\end{flushright}

	\begin{teo}
		\label{Resolvent2}
		For each \( \alpha \in \mathrm{C}(S) \), \( \alpha(r) \neq 0 \), \( \forall \: r \in S \), the sequence of semibounded self-adjoint operators \( K_{\alpha}^N \) converge as \( N \rightarrow \infty \) in the strong resolvent sense to the self-adjoint (unbounded from below) operator \( K_{\alpha} \):
		\bdm
			\mathcal{D}(K_{\alpha}) = \big\{ \Psi \in L^2(\mathbb{R}^2) \: | \: \exists \xi_{\Psi} \in L^2(S, d\mu_S), \Psi - \tilde{\mathcal{G}}_z \xi_{\Psi} \in \mathcal{D}(H_{\omega}),
		\edm
		\martin
			\big( \Psi - \tilde{\mathcal{G}}_z \xi_{\Psi} \big)\big|_S = \Gamma_{\alpha}(z) \xi_{\Psi} \big\}
		\sileno
		\martin
		\label{Ham4}
			\big( K_{\alpha} -z \big) \Psi = \big( H_{\omega} - z \big) \big( \Psi - \tilde{\mathcal{G}}_z \xi_{\Psi} \big)
		\sileno
		where
		\martin
			\label{Gam3}
			\Big[ \Gamma_{\alpha}(z) \: \xi_{\Psi} \Big] (r) = \frac{\xi_{\Psi}(r)}{\alpha(r)} - \int_S d\mu_S(r^{\prime}) \:\: \mathcal{G}_{z}(\vec{x}, \vec{x}^{\prime})\big|_{\vec{x}, \vec{x}^{\prime} \in  S} \: \xi_{\Psi}(r^{\prime})
		\sileno
		\bdm
			\big(\tilde{\mathcal{G}}_{z} \xi \big) (\vec{x}) \equiv \int_S d\mu_S(r^{\prime}) \:\: \mathcal{G}_z(\vec{x}, \vec{x}^{\prime})\big|_{\vec{x}^{\prime} \in D} \: \xi(r^{\prime})
		\edm
		Moreover the resolvent of \( K_{\alpha} \) is			
		\bdm
			\Big[ \big( K_{\alpha} - z \big)^{-1} \Psi \Big](\vec{x}) = \Big[ \big( H_{\omega} - z \big)^{-1} \Psi \Big] (\vec{x}) \: + 
		\edm
		\martin
			\label{Resolvent4}
			+ \int_S dr^{\prime} \: r^{\prime} \:  \Gamma^{-1}_{\alpha}(z) \Big[ \big( H_{\omega} - z \big)^{-1} \Psi \Big]\Big|_S (r^{\prime}) \: \mathcal{G}_z(\vec{x}, \vec{x}^{\prime})\big|_{\vec{x}^{\prime} \in S}
		\sileno
		for each \( z \in\varrho(K_{\alpha}) \).
	\end{teo}

	\emph{Proof:}	See the Proof of Theorem \ref{Resolvent1}.
		\begin{flushright} 
			\( \Box \)
		\end{flushright}

	\begin{teo}
		\label{Spe4}
		The spectrum of \( K_{\alpha} \) is purely absolutely continuous and
		\bdm
			\sigma(K_{\alpha}) = \sigma_{\mathrm{ac}} (K_{\alpha}) = \sigma(H_{\omega}) = \mathbb{R}
		\edm
	\end{teo}
	
	\emph{Proof:}
		See the Proof of Theorem \ref{Spe3}, Theorem \ref{Resolvent2} and Proposition \ref{Gre2}.
		\begin{flushright} 
			\( \Box \)
		\end{flushright}
	
\textbf{Remark.}	An interesting application of previous results is the study of the 3D rotating needle, i.e. a singular rotating perturbation of the Laplacian supported on a (finite) segment. Indeed the system can be easily reduced to a 2D rotating blade on the plane of rotation and a free motion on its perpendicular axis: the Hamiltonian is formally given by
	\bdm 
		H = H_0^{x,y} + \alpha(x) \: R(t) \: \Theta_A(x) \: \delta(y) + H_0^{z}
	\edm	
where \( \Theta_A(x) \) is the characteristic function of the segment \( 0 \leq x \leq A \). According to the previous discussion, the self-adjoint extensions of \( H \) are given by the family of operators \( K_{\alpha}^{x,y} + H_0^z \), where \( K_{\alpha}^{x,y} \) denotes the Hamiltonians of the 2D rotating blade defined in (\ref{Ham4}). Moreover the domain of self-adjointness can be identified with the set of functions \( \Psi(\vec{x}) = f(x,y) \: g(z) \) such that \( f \in \mathcal{D}(K_{\alpha}) \) and \( g \in H^2(\mathbb{R}) \).

\subsection{Asymptotic Limit of Rapid Rotation}

In this Section, we shall prove that
	\bdm
		\mathrm{s-}\lim_{\omega \rightarrow \infty} U_{\mathrm{inert}}(t,s) = e^{-iH_{\alpha} (t-s)}
	\edm
where \( H_{\alpha} \) is the self-adjoint generator
	\martin
		\label{needle2}
		H_{\alpha} = H_0 - \alpha(r) \: \Theta_{C}(r)
	\sileno	 
and \(  \Theta_C(r) \) is the characteristic function of a circle \( C \) of radius \( A \) centered at the origin. 
	
	\begin{teo}
		\label{needle1}
		For every \( t,s \in \mathbb{R} \),
		\bdm
			\mathrm{s-}\lim_{\omega \rightarrow \infty} U_{\mathrm{inert}}(t,s) = e^{-iH_{\alpha} (t-s)}
		\edm
		where 
		\bdm
			H_{\alpha} = H_0 - \alpha(r) \: \Theta_{C}(r)
		\edm
	\end{teo}

	\emph{Proof:}	See the Proof of Theorem \ref{Asy1} and the following Lemma \ref{Con4}.
		\begin{flushright} 
			\( \Box \)
		\end{flushright}

	\begin{lem}
		\label{Con4}
		For every \( z \in \mathbb{C} \), \( \Im(z) > 0 \),
		\bdm
			\mathrm{s-}\lim_{\omega \rightarrow \infty} \int_{-\infty}^0 dt \: e^{-izt} \: U^{*}_{\mathrm{inert}}(t,0) = -i \big(H_{\alpha} - z \big)^{-1}
		\edm
	\end{lem}
		
	\emph{Proof:}	Like in the Proof of Lemma \ref{Con2}, we shall prove the result on the dense subset of \( L^2(\mathbb{R}^2) \) given by functions of the form \( \Psi(\vec{x}) = \chi(r) e_{n_0}(\theta) \), \( n_0 \in \mathbb{Z} \). Following the Proof of Lemma \ref{Con2}, it remains to prove that
		\bdm
			\lim_{\omega \rightarrow \infty} \big( K_{\alpha}+ n_0 \omega - z \big)^{-1} \Psi(\vec{x}) =  \big(H_{\alpha} - z \big)^{-1} \Psi(\vec{x})
		\edm
		but
		\bdm
			 \big( K_{\alpha} + n_0 \omega - z \big)^{-1} \Psi = \big( H_{\omega} + n_0 \omega - z \big)^{-1} \Psi \: +
		\edm
		\bdm
			+ \: \Big(  \Gamma^{-1}_{\alpha}(z^* - n_0 \omega) \Big[ \big( H_{\omega} + n_0 \omega - z^* \big)^{-1} \Psi \Big]\Big|_S \: , \:  \mathcal{G}_{z - n_0 \omega}(\vec{x}, \vec{x}^{\prime})\big|_{\vec{x}^{\prime} \in S} \Big)_{L^2(S, d\mu_S)}
		\edm
		and
		\bdm
			\lim_{\omega \rightarrow \infty} \big( H_{\omega} + n_0 \omega - z \big)^{-1} \Psi = \big(H_0 - z \big)^{-1} \Psi
		\edm
		as we have proved in Lemma \ref{Con2}. Moreover 
		\bdm
			\lim_{\omega \rightarrow \infty} \Big[ \big( H_{\omega} + n_0 \omega - z \big)^{-1} \Psi \Big]\Big|_S = \Big[ \big(H_0 - z \big)^{-1} \Psi \Big]\Big|_S
		\edm
		in \( L^2(S, d\mu_S) \) and, applying the result found in the following Lemma \ref{Gamma2},
		\bdm
			\lim_{\omega \rightarrow \infty} \Gamma^{-1}_{\alpha}(z - n_0 \omega) \Big[ \big( H_{\omega} + n_0 \omega - z \big)^{-1} \Psi \Big]\Big|_S = 
		\edm
		\bdm
			= \Xi_{\alpha}(z) \Big[ \big( H_0 - z \big)^{-1} \Psi \Big]\Big|_S =
		\edm
		\bdm
			= \alpha(r) \Theta_S(r) \big( H_0 - \alpha(r) \: \Theta_C(r) - z \big)^{-1} \Psi
		\edm
		In conclusion we obtain
		\bdm
			\lim_{\omega \rightarrow \infty} \big( K_{\alpha}+n_0 \omega - z \big)^{-1} \Psi = \big( H_0 - z \big)^{-1} \Big[ 1 + \alpha \Theta_S \big( H_0 - \alpha \: \Theta_C - z \big)^{-1} \Big] \Psi =
		\edm
		\bdm
			= \big( H_0 - \alpha \: \Theta_C - z \big)^{-1} \Psi
		\edm
		\begin{flushright} 
			\( \Box \)
		\end{flushright}

	\begin{lem}
		\label{Gamma2}
		Let \( \Gamma_{\alpha}(z) \) the operator defined in (\ref{Gam3}),
		\bdm
			\lim_{\omega \rightarrow \infty} \Gamma^{-1}_{\alpha}(z - n_0 \omega) = \Xi_{\alpha}(z) 
		\edm
		in \( L^2(S, d\mu_S) \), where
		\martin
			\big( \Xi_{\alpha}(z) \xi \big) (r) \equiv \bigg[ \alpha(r) \Big( H_0 - \alpha(r) \: \Theta_C(r) - z \Big)^{-1} \big( H_0 -z \big) \: \xi \bigg] (r)
		\sileno
	\end{lem}

	\emph{Proof:}	First of all we are going to prove that
		\bdm
			\mathrm{norm-}\lim_{\omega \rightarrow \infty} \Gamma_{\alpha}(z - n_0 \omega) = \Lambda_{\alpha}(z)
		\edm
		where
		\bdm
			\Lambda_{\alpha}(z) \: \xi = \frac{\xi}{\alpha} - \frac{1}{2 \pi} \int_S d\mu_S(r^{\prime}) \:\: g_z^{n_0}(r, r^{\prime}) \xi(r^{\prime})
		\edm
		for the definition of \( g_z^{n_0} \) see Proposition \ref{ReH3}.
		\newline
		Indeed
		\bdm
			\Gamma_{\alpha}(z - n_0\omega) = \Lambda_{\alpha}(z) + R_z^{n_0}
		\edm
		where \( R^{n_0}_z \) is a bounded integral operator on \( L^2(S, d\mu_S) \) with kernel
		\bdm
			R^{n_0}_z(r, r^{\prime}) \equiv \int_0^{\infty} \underset{n \neq n_0}{\sum_{n=-\infty}^{\infty}} \frac{\varphi_{kn}(r) \: \varphi_{kn}(r^{\prime})}{k^2 - (n - n_0) \omega - z} \longrightarrow 0
		\edm
		as \( \omega \rightarrow \infty \) (see the Proof of Lemma \ref{Con2}).
		\newline
		Moreover for each \( n_0 \in \mathbb{Z} \) and \( z \in \mathbb{C} - \mathbb{R} \) it can be seen that the operator \( \Lambda_{\alpha} \) is invertible: indeed
		\bdm
			\Big[ \big(H_0 - z \big) \Lambda_{\alpha}(z) \: \xi \Big] (\vec{r}) = \bigg[ \big(H_0 - z \big) \frac{\xi}{\alpha} \bigg] (r) - \Theta_S(r) \: \xi(r)
		\edm
		so that \( \Lambda^{-1}_{\alpha}(z) = \Xi_{\alpha}(z) \). 
		\begin{flushright} 
			\( \Box \)
		\end{flushright}

\textbf{Remark.}	As in Section 5.1, we can apply the previous results to analyse the asymptotic limit of rapid rotation of the 3D rotating needle: the time-dependent propagator in the inertial frame factorizes in the product of the time-dependent propagator associated to a 2D rotating blade on the \(x,y\)-plane and a the free propagator on the \(z\)-axis. Therefore Theorem \ref{needle1} implies convergence of the propagator in the inertial frame to the one-parameter unitary group generated by the time-independent self-adjoint operator \( H_{\alpha}^{x,y} + H_0^z \), where \( H_{\alpha}^{x,y} \) is defined in (\ref{needle2}).

\section{Conclusions and Perspectives}

The operators studied in Section 2 and 3 could be viewed as the Hamiltonians of quantum systems given by a particle interacting with a rotating \(\delta\)-type potential. In this context the results proved about the asymptotic limit of rapid rotation have an heuristic physical meaning: if the angular velocity of the potential is very large with respect to the velocity of the particle, we expect that the particle feels a time-independent potential, which is the mean of the true potential over a period.
\newline
This result was already proved by Enss et al. \cite{Enss2} for regular potential, and, from this point of view, our work is an extension of their results to singular potentials.
\newline
A future application of that study would be the analysis of the scattering of a particle by a rotating point interaction. Indeed it would be an example of time-dependent scattering that can be reduced to a stationary problem: passing to the rotating frame, we could prove in simpler way, for example, existence and completeness of the wave operators.
\newline
In Section 3 and 4 we have studied the rotating blade, namely a singular potential with codimension 1. That kind of rotating singular perturbations of the Laplacian are more interesting and could open many suggestive problems. 
\newline
For example in the 3D case we could investigate the dependence of the results on the shape of the blade. While all the properties of the form and the self-adjoint extensions still hold for a blade with a general shape, because the key point is the good behavior of the Green's function on a compact subset of \( \mathbb{R}^3 \), the analysis of the asymptotic limit is harder. 
\newline
In fact a semi-spherical shape is very useful to perform the calculation with the Green's function of \( H_{\omega} \) expressed in terms of functions with spherical symmetry (the spherical waves), but the same goal can be reached for a blade of different form: if we take a square shaped blade and we express the resolvent of \( H_{\omega} \) in terms of functions with cylindrical symmetry (essentially the Bessel functions), all the results still hold. On the other hand, if the blade has no symmetry, we could expect the same behavior but it is not clear at all how it can be proved. 
\newline
Finally we want to mention another feature of the problem which can be investigated: the blades we have considered are finite, so it would be interesting to study an infinite blade, for example an half-line in 2D and an half-plane in 3D, but, in that case, many problems arise in the definition of the operator. In particular the form \( \Phi_{\alpha}^z \) should not be bounded, unless we impose some condition on the behavior at \( \infty \) of the parameter \( \alpha \). 
\newline
\mbox{}	\\
\mbox{}	\\
\textbf{Acknowledgments:} M.C. is very grateful to Prof. Ludwik Dabrowski  and the INTAS Research Project nr. 00-257 of European Community, ``Spectral Problems for Schr\"{o}dinger-Type Operators'', for the support.

\newpage

\appendix

\section*{APPENDIX}

\section{The Green's Function of \( H_{\omega} \)}

In this Appendix we shall study the Green's function \( \mathcal{G}_z (\vec{x}, \vec{y}_0) \) of \( H_{\omega} \) and we shall prove that it belongs to \( L^2(\mathbb{R}^n, d^n\vec{x}) \), \( \forall \vec{y}_0 \in \mathbb{R}^n \) with \( n = 2,3 \).
\newline
We shall start from the 3D case:

	\begin{pro}
		\label{Gre1}
		The resolvent \( (H_{\omega} - z)^{-1} \), \( z \in \mathbb{C}-\mathbb{R} \), has the following integral representation
		\bdm
	 		(H_{\omega} - z)^{-1} \Psi(\vec{x}) = \int_{\mathbb{R}^3} d^3x' \mathcal{G}_z(\vec{x}, \vec{x}') \Psi(\vec{x}')
		\edm
		with \( \Psi (\vec{x}) \in L^2(\mathbb{R}^3, d^3 x) \) and
		\martin
			\mathcal{G}_z(\vec{x},\vec{x}') = \int_0^{\infty} dk \sum_{l=0}^{\infty} \sum_{m = -l}^{l} \frac{1}{k^2 - m \omega - z} \: \varphi^*_{klm} (\vec{x}') \: \varphi_{klm}(\vec{x})
		\sileno
		The functions \( \varphi_{klm}(\vec{x}) \) are the spherical waves\footnote{Here \( j_l(r) \) denotes the spherical Bessel function of order \( l \) (see \cite{Niki1, Wats1}) and \( Y_l^m(\theta, \phi) \), with \( l \in \mathbb{N} \) and \( m = -l, \ldots, l \), the spherical harmonics.}:
		\bdm
			\varphi_{klm}(\vec{x}) = \sqrt{\frac{2k^2}{\pi}} j_l(kr) Y^m_l(\theta, \varphi)
		\edm
		Moreover, for every \( \vec{y}_0 \in \mathbb{R}^3 \) and \( z \in \mathbb{C}-\mathbb{R} \), \( \mathcal{G}_z(\vec{x},\vec{y}_0) \in L^2(\mathbb{R}^3, d^3\vec{x}) \).
	\end{pro}
		
	\emph{Proof:}	The integral representation of the Green's function of \( H_{\omega} \) is a straightforward consequence of the eigenvectors decomposition of \( H_{\omega} \). Moreover in the following we shall prove that, for each \( \Psi \in L^2(\mathbb{R}^3) \), \( z \in \mathbb{C}-\mathbb{R} \) and \( \vec{y}_0 \in \mathbb{R}^3 \),
		\bdm
			\Big| \Big( \mathcal{G}_{z}(\vec{x}, \vec{y}_0) \: , \: \Psi(\vec{x}) \Big)_{L^2(\mathbb{R}^3, d^3\vec{x})} \Big| < \infty 
		\edm
		Every function \( \Psi \in L^2(\mathbb{R}^3) \) can be decomposed in terms of spherical harmonics:
		\bdm
			\Psi(\vec{x}) = \sum_{l=0}^{\infty} \sum_{m=-l}^l \: \Psi_{lm}(r) \: Y_l^m(\theta, \phi) 
		\edm
		with the \(L^2\)-condition 
		\bdm	
			\sum_{l=0}^{\infty} \sum_{m=-l}^l \: \big\| \Psi_{lm}(r) \big\|^2_{L^2(\mathbb{R}^+, r^2dr)} < \infty
		\edm
		Thus
		\bdm
			\Big| \Big( \mathcal{G}_{z}(\vec{x}, \vec{y}_0) \: , \: \Psi(\vec{x}) \Big) \Big|^2 \leq \sum_{l=0}^{\infty} \sum_{m=-l}^l \Big| \Big( G_{z+m\omega}(\vec{x}, \vec{y}_0) \: , \: \Psi_{lm}(r) Y_l^m(\theta, \phi) \Big) \Big|^2 \leq
		\edm
		\bdm
			\leq \sum_{l=0}^{\infty} \sum_{m=-l}^l \big\| G_{z+m\omega}(\vec{x}, \vec{y}_0) \big\|^2_{L^2(\mathbb{R}^3, d^3\vec{x})} \big\| \Psi_{lm}(r) Y_l^m(\theta, \phi) \big\|^2 \leq
		\edm
		\bdm
			\leq C(\Im(z)) \: \sum_{l=0}^{\infty} \sum_{m=-l}^l \: \big\| \Psi_{lm}(r) \big\|^2_{L^2(\mathbb{R}^+, r^2dr)} < \infty
		\edm
	 	because the Green's function of the free Hamiltonian
		\bdm	
			G_{z+m \omega} (\vec{x}, \vec{x}) = \frac{e^{i\sqrt{z + m\omega}|\vec{x}- \vec{y}_0|}}{4 \pi |\vec{x} - \vec{x}|}
		\edm
		belongs to \( L^2(\mathbb{R}^3, d^3\vec{x}) \) for each \( z \in \mathbb{C}-\mathbb{R} \) and \( \vec{y}_0 \in \mathbb{R}^3 \): we have to choose the root of \( z + m \omega \) with imaginary part 
		\bdm
			\Im \big( \sqrt{z + m\omega} \big) = \sqrt{ \frac{ \Big[ (\Re(z)+m\omega)^2 + \Im(z)^2 \Big]^{\frac{1}{2}} - \Re(z) - m \omega}{2}} \geq \sqrt{\frac{|\Im(z)|}{2}} > 0
		\edm
		so that \( G_{z+m \omega} \in L^2 \) independently on \( m \in \mathbb{Z} \).
		\begin{flushright} 
			\( \Box \)
		\end{flushright}

An analogous result can be proved in the 2D case:

	\begin{pro}
		\label{Gre2}
		The resolvent \( (H_{\omega} - z)^{-1} \), \( z \in \mathbb{C}-\mathbb{R} \), has the following integral representation
		\bdm
	 		(H_{\omega} - z)^{-1} \Psi(\vec{x}) = \int_{\mathbb{R}^2} d^2x' \mathcal{G}_z(\vec{x}, \vec{x}') \Psi(\vec{x}')
		\edm
		with \( \Psi (\vec{x}) \in L^2(\mathbb{R}^2, d^2 x) \) and\footnote{ \( J_{n}(r) \) stands for the Bessel function of order \( n \in \mathbb{N} \).}
		\martin
			\mathcal{G}_z(\vec{x},\vec{x}') \equiv \int_0^{\infty} dk \sum_{n=-\infty}^{\infty} \frac{1}{k^2 - \omega n - z} \: \varphi^*_{kn} (\vec{x}') \: \varphi_{kn}(\vec{x})
		\sileno
		\bdm
			\varphi_{kn}(\vec{x}) = \sqrt{\frac{k}{2 \pi}} J_{|n|}(kr) \: e^{in \theta}
		\edm
		Moreover, for every \( \vec{y}_0 \in \mathbb{R}^2 \) and \( z \in \mathbb{C}-\mathbb{R} \), \( \mathcal{G}_z(\vec{x},\vec{y}_0) \in L^2(\mathbb{R}^2, d^2\vec{x}) \).
	\end{pro}
		
	\emph{Proof:}	Following the Proof of Proposition \( \ref{Gre1} \), we shall consider the scalar product
		\bdm
			\Big( \mathcal{G}_{z}(\vec{x}, \vec{y}_0) \: , \: \Psi(\vec{x}) \Big)_{L^2(\mathbb{R}^3, d^3\vec{x})}
		\edm
		with 
		\bdm
			\Psi(\vec{x}) = \sum_{n=-\infty}^{\infty} \Psi_n(r) \: \frac{e^{in\theta}}{2\pi} 
		\edm
		and we obtain
		\bdm
			\Big| \Big( \mathcal{G}_{z}(\vec{x}, \vec{y}_0) \: , \: \Psi(\vec{x}) \Big) \Big|^2 \leq  \sum_{n=-\infty}^{\infty} \big\| G_{z+m\omega}(\vec{x}, \vec{y}_0) \big\|^2_{L^2(\mathbb{R}^3, d^3\vec{x})} \big\| \Psi_{n}(r) \big\|^2_{L^2(\mathbb{R}^+, r^2dr)} < \infty
		\edm
		since\footnote{\( H_0^{(1)} \) denotes the Hankel function of first kind and order zero (see \cite{Abra1}).} 
		\bdm
			G_{z+n \omega} (\vec{x}, \vec{y}_0) = \frac{i}{4} \: H_0^{(1)}(\sqrt{z+n\omega} \: |\vec{x}- \vec{y}_0|)
		\edm
		belongs to \( L^2(\mathbb{R}^2, d^2\vec{x}) \), for each \( z \in \mathbb{C}-\mathbb{R} \) and \( \Im(\sqrt{z+n\omega}) > 0 \).
		\begin{flushright} 
			\( \Box \)
		\end{flushright}

\end{document}